# High-Efficiency Deep Blue Single-Gaussian Europium(II) Emitters and their Emitter-Host Interactions

*Mahmoud Soleimani,* [a,b] *Paulius Imbrasas,* [b] *Jan-Michael Mewes,* [b]* *Felix Kaden,* [b] *Stephanie Anna Buchholtz,* [a] *Karl Leo,* [a] *Sebastian Schellhammer,* [a] *Carsten Rothe,* [b] *Sebastian Reineke* [a]*

[a] Institute of Applied Physics (IAP) and Dresden Integrated Center for Applied Physics and Photonic Materials (IAPP), Technische Universität Dresden, Dresden, Germany
[b] beeOLED GmbH, Dresden, Germany

DFT correspondence E-mail: jan.mewes@beeoled.com
Experiments correspondence E-mail: sebastian.reineke@tu-dresden.de

Funding and acknowledgments: Financial support from the project "Blaue Emitter" (SAB no. 100671218 and 100671145) is gratefully acknowledged. The project is co-funded by the European Union and co-financed from tax revenues on the basis of the budget adopted by the Saxon State Parliament.

## Abstract

Eu(II) complexes are attractive emitters for deep-blue organic light-emitting diodes (OLEDs) due to their narrow, parity-allowed 4*f*–5*d* emission; however, their implementation in vacuum-processed OLEDs has remained limited. Here, we introduce a new molecular design concept for Eu(II) emitters, in which a crown-ether ligand is combined with carborate anions to define the coordination environment and improve steric shielding of the europium center. Based on this design, we present two emitters that combine narrow deep-blue photoluminescence with quantum yields approaching 90% and sufficient thermal stability for vacuum deposition. As the excited state dynamics of this emitter class are different from most conventional OLED emitters and the pathway to maximum luminescence efficiency in thin films is not fully established, we study interactions between Eu(II) complexes and the host environment, based on density functional theory and time-resolved experiments. We identify steric shielding of the Eu(II) core and energetic confinement of the excited 5*d* electron, defined by molecular design as key factors governing efficient luminescence, providing a roadmap for rational design of Eu(II) emitters. Together, these results establish a basis for higher-efficiency and deeper-blue OLEDs incorporating Eu(II) emitters.

## 1 Introduction

Organic light-emitting diodes (OLEDs) are widely used in small and medium-sized displays, such as smartwatches and smartphones.[1] However, the use of this technology in large-scale applications is hindered by the performance limitations of blue OLEDs.[2] While red and green OLEDs have met industrial requirements for full commercialization, blue OLEDs continue to struggle to combine high efficiency and long-term stability.[3]

The shorter lifetime of highly efficient blue emitters in OLEDs compared to their red and green counterparts is in part due to the higher energy of the molecular excitations leading to blue light emission.[3] Many state-of-the-art blue-emitting materials with high efficiency are based on metal-organic structures, which undergo organic bond cleavage during device operation.[4] In contrast, purely organic fluorescent emitters are generally stable but suffer from intrinsically low efficiency due to their inability to harvest triplet excitons. Moreover, triplet-harvesting emitters based on thermally activated delayed fluorescence (TADF) can, in principle, achieve near-unity exciton utilization owing to efficient intersystem crossing. However, despite their high efficiencies, TADF emitters often face challenges in long-term operational stability similar to many transition-metal-organic counterparts. Lacking a satisfying pathway to blue OLEDs that are on par with their respective green and red counterparts, material development efforts to find high-performing blue emitter materials are still high.[3] Currently, two main emitter classes are widely studied and developed: phosphorescent emitters based on Pt(II) and multi-resonant (MR-)TADF emitters based on a "DABNA-core".[5] Further but less prominent approaches are doublet emitters whose spin-state can interact with singlets and triplets,[6] and emitters with an inverted singlet-triplet gap.[7]

Lanthanide-based emitters are another promising class that combines heavy-element effects with spin states allowing interaction with both singlet and triplet excitons. For blue emission, both Ce(III) and Eu(II) exhibit parity-allowed 4*f*–5$d$ transitions in a suitable energy range, enabling, in principle, 100% exciton utilization.[8] Ce(III) has a doublet ground state and exhibits double-peaked emission, which makes it less suitable as a terminal emitter; however, owing to its short excited-state lifetime, it is more suitable as a sensitizer.[9] In contrast, Eu(II) has an octet $4f^7$ ($^8S_{7/2}$) ground state with zero orbital angular momentum, resulting in a single, narrow $5d \rightarrow 4f$ emission band with high color purity, making it well-suited as a terminal emitter for deep-blue emission.

Over the past decades, numerous studies have focused on Eu-based emitters, with early work dating back to 1999, when Shipley et al. introduced an orange-emitting divalent europium complex based on borate ligands; however, the corresponding electroluminescent device exhibited very poor performance.[10] In 2018, Allen and co-workers reported an azacryptand-based Eu(II) complex showing highly efficient emission with a photoluminescence quantum yield (PLQY) of up to 100% and a narrow deep-blue emission peak at 450 nm. Nevertheless, this complex exhibited salt-like properties, resulting in low volatility and solubility only in highly polar solvents, thereby preventing its use in state-of-the-art OLED fabrication.[11] Here, volatility refers to the ability of a material to be sublimed under high vacuum without decomposition, which is essential for vacuum deposition processes. In 2020, Liu et al. reported a related azacryptand complex (EuCrypt, **Figure 2**), which showed improved sublimation behavior and enabled OLED fabrication with external quantum efficiencies (EQEs) of up to 17.7%; however, emission was limited to the green spectral region.[12] Due to its specific electronic structure, EuCrypt exhibits a broad double-band emission around 560 nm rather than the desired single-band deep-blue emission. In the same year, Liu and co-workers also reported borate-based Eu(II) complexes with orange and sky-blue emission, with the sky-blue emitter showing EQEs below 1%.[13] Subsequent work by Liu in 2023 demonstrated a deep-blue-emitting Eu complex; however, devices fabricated via solution processing, not vacuum deposition, exhibited limited performance, with maximum EQEs of only 4.7%.[14] More recently, Liu et al. (2024) reported Eu complexes based on bulky tert-butylated tris(pyrazolyl)borate ligands, achieving improved sublimation behavior. The best device reached a maximum EQE of 15.7%, yet still showed sky-blue emission with a peak wavelength of 478 nm.[15] Despite these advances, to the best of our knowledge, the combination of deep-blue emission with high color purity, sufficient thermal stability, and volatility compatible with vacuum deposition has not yet been achieved within a single Eu(II) system.

In this work, we address two central points: i) We introduce two new Eu(II) complexes that are developed to retain the Eu(II) configuration through confinement that is essential for blue emission and to enable thermal evaporation under high vacuum, a prerequisite for industry-scale fabrication. The newly developed Eu(II) emitters combine intense single-band deep-blue emission with photoluminescence quantum yields of up to 90% and long excited-state lifetimes, confirming highly efficient 4*f*–5*d* emission. Importantly, both emitters are compatible with vacuum processing, exhibiting thermal stability with high sublimation yields, enabling thermal evaporation in OLED fabrication tools. We demonstrate their processability through the fabrication of OLEDs based on a representative reference architecture using a standard thermal evaporation tool, serving as a proof of concept for device integration (for details, *cf.* Supporting Information, Section S4). The two emitters were judiciously selected as a related yet distinct molecular pair, providing a unique comparative platform to study how differences in steric shielding and excited-state energetics govern host compatibility and luminescence efficiency. ii) We discuss the photoluminescence of these emitters when embedded in different host materials that all have exothermic optical bandgaps compared with the Eu(II) transition, but differ in their energetic position of the frontier orbitals. This latter study identifies an emitter-specific quenching channel through coordination and electron transfer that must be avoided in the design of OLEDs incorporating these emitters. Although Eu(II) complexes can exhibit excellent intrinsic photophysical properties, their translation into efficient and color-pure devices is governed by excited-state confinement and effective shielding of the Eu(II) center. By combining photoluminescence studies in evaporated thin films with density functional theory (DFT)–based electronic-structure calculations, we establish a framework that directly links molecular shielding, excited-state energetics, and host compatibility. These insights explain the previously unattained combination of deep-blue emission, high color purity, and vacuum-deposition compatibility achieved here, as well as the limitations of previous Eu(II)-based emitters and provide design principles for future Eu(II) emitters with optimal electroluminescent efficiency.

**Electronic Structure of Eu(II) Emitters**

Eu(II) emitters make use of parity-allowed $4f$–$5d$ transitions, which in principle enable 100% exciton utilization, and intrinsically single-band emission leading to high color purity. Compared to conventional $\pi$-conjugated molecules typically employed in OLEDs, Eu(II) emitters have a fundamentally different electronic structure, which needs to be understood to enable their efficient implementation in OLEDs. Eu(II) features a half-filled $4f^7$ ground state, and its emission does not originate from a HOMO–LUMO (highest occupied and lowest unoccupied molecular orbital) transition. As a consequence, the energy levels governing charge injection into the emitter, excited-state confinement, and charge recombination are defined by the metal-centered electronic structure and its interaction with the surrounding environment, rather than by molecular frontier orbitals.

In the ground state, Eu(II) features a half-filled $4f^7$ shell, while the diffuse $5d$ manifold lies above the vacuum level and therefore does not constitute a bound state (**Figure 1**). Upon excitation, the creation of a $4f$ hole substantially stabilizes the $5d$ levels, binding the excited electron at the Eu center. As a consequence, the energetic position of the excited electron strongly depends on the oxidation state of europium, and the conventional LUMO estimation of $\pi$-conjugated emitters is no longer applicable.

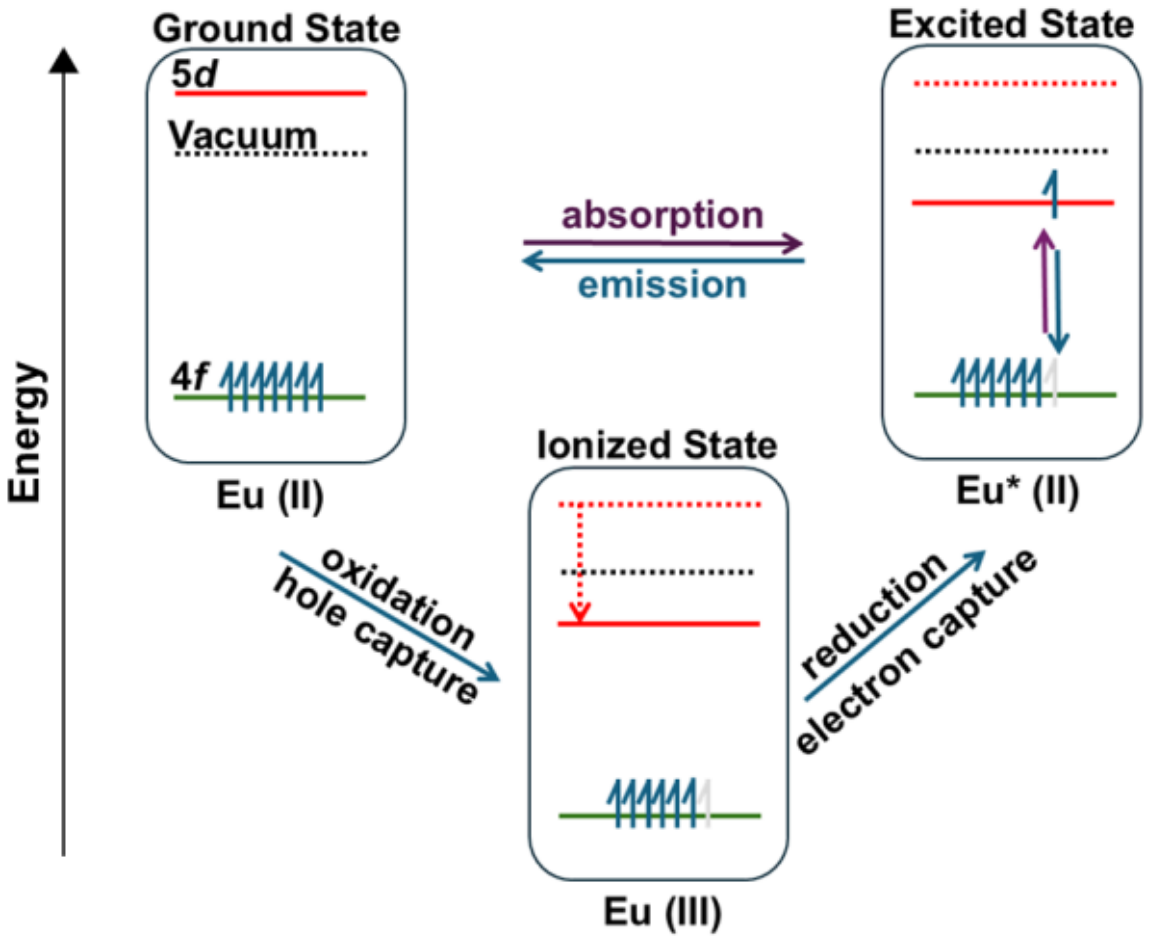

**Figure 1**. Electronic Structure of Eu(II) Emitter. Energy levels and population of the 5*d* and 4*f* shells of Eu(II) in the ground state, excited state, and oxidized state [Eu(III)].

A characteristic feature of Eu(II) complexes is their susceptibility to oxidation to the trivalent state Eu(III), as documented in numerous stability studies.[16,17] The shallow energetic position of the half-filled 4*f* level renders Eu(II) prone to electron loss under chemical or electrical perturbation. In the context of light emission, transient conversion to Eu(III) is required to generate $d - f$ excited Eu(II). However, permanent chemical conversion into Eu(III) eliminates the parity-allowed 4*f*–5*d* transition and the desired blue emission. The stability of divalent europium is not solely determined by the metal center itself but is influenced by its local chemical and electrostatic environment. Avoiding permanent oxidation to trivalent europium is essential for stable and efficient Eu(II)-based electroluminescence.

The influence of the local environment on the stability of Eu(II) can be rationalized using the hard–soft acid–base (HSAB) principle.[18] As a comparatively soft Lewis acid, Eu(II) is preferentially stabilized by soft, weakly coordinating donors and counter-ions. In contrast, hard or strongly coordinating Lewis bases, such as heteroatoms commonly present in OLED host materials, can promote oxidation to Eu(III). Beyond chemical stability, efficient emission from Eu(II) emitters requires confinement of the excited electron on the metal center during exciton formation, such that oxidation to Eu(III) and excited-state electron loss represent closely related manifestations of insufficient electron confinement. This charge confinement can be evaluated using excited-state ionization energy (ES–IE), which serves as a predictor for excited-electron transfer to acceptor states of the host. When the ES–IE lies above the host LUMO, electron confinement is weak, facilitating electron transfer and emission quenching, whereas an ES–IE below the host LUMO indicates effective confinement and efficient luminescence. However, quenching pathways additionally require sufficient orbital overlap between the Eu 5*d* state and host acceptor orbitals, which can be suppressed by effective steric and electronic shielding of the Eu(II) center. Consequently, both the choice of ligands and the nature of the counter-ions play a decisive role in controlling the electronic structure, stability, and device compatibility of Eu(II) emitters. Clearly, in particular with respect to their use in OLEDs, the overall understanding of the excitation dynamics is limited and needs a deeper understanding to make sure that electroluminescence can be obtained from undisturbed atomic transitions in the Eu(II) center. Based on these principles, we further investigate the Eu(II) emitters.

## 2 Results and Discussion

### 2.1 Material Design and Synthesis

EuCrypt is a Eu(II)-based emitter that was previously employed in OLEDs by Liu et al. in 2020 and is included for reference.[12] In EuCrypt, Eu(II) is coordinated by six $HNR_2$ and two $NR_3$ donors of an aza-cryptand and neutralized by two iodide anions, one of which is coordinating Eu (r(Eu-I)=3.65 Å), the other one in the outer coordination sphere bound to NH protons (r(Eu-I)=3.95 Å, *cf.* **Figure 2**). The DFT results obtained at the ωB97X-D3 level of theory are in reasonable agreement with the X-ray structure reported by Liu.[12] The reported emission spectrum of EuCrypt is atypically broad for Eu(II) and depends on the chemical environment, covering the green to yellow spectral range depending on the solvent.[12]

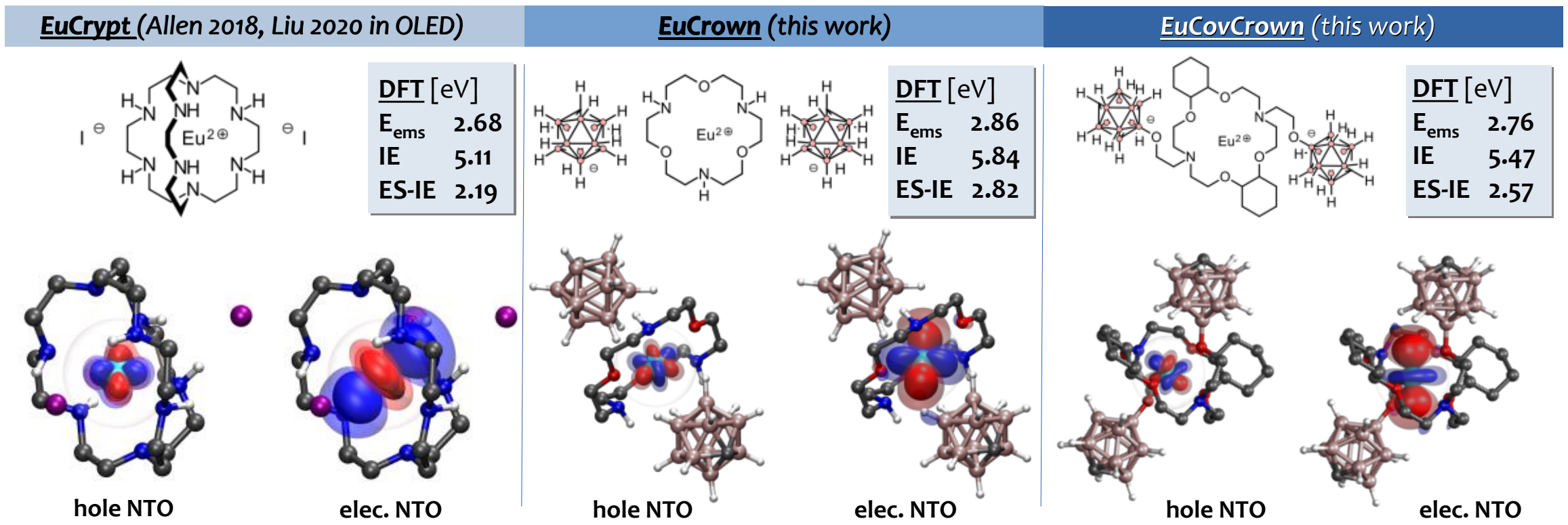

**Figure 2.** Molecular design and electronic states of Eu(II) emitters studied in this work. 2D and 3D structures, along with hole and electron natural transition orbitals (NTOs) of the lowest lying excited state, were calculated at the TD-DFT/$\omega$B97X-D3/SMD/LR-PCM level of theory for the excited-state optimized structures. Shown is the first NTO-pair ($4f$-hole on the left, $5d$-electron on the right), which makes up ≈ 99.5% of the transition for all emitters. Visualization using VMD 1.9.2[19] with isosurfaces for 0.94e (solid) and 0.98e (transparent). For clarity, all hydrogen atoms bonded to carbon are hidden.

For EuCrypt, the DFT analysis (*cf.* Methods) yields a high dipole moment of 12.7 D, an ionization energy (IE = −HOMO) of 5.1 eV, and an excited-state (ES)−IE of 2.2 eV (−LUMO level in an organic emitter). These values are much shallower than the HOMO and LUMO levels of −5.7 eV and −3.3 eV reported by Liu from Ultraviolet photoelectron spectroscopy (UPS) measurements, but they agree better with the behavior observed in thin films here and elsewhere.[12] With an ES−IE of 2.2 eV, EuCrypt starts to lose electron confinement in hosts with a LUMO below −2.2 eV, excluding all but the shallowest host materials. For example, ref. [12], reported the best performance for m-MTDATA (LUMO of −2.0 eV) in the emissive layer, whereas deeper hosts showed no or strongly reduced emission. In addition, such a large dipole moment is consistent with strong electrostatic interactions in the solid state and thus with the limited volatility and salt-like character of the material. Further discussion of the dipole moment and the sensitivity of the emission energy to molecular conformation is given in Section S6 of the Supporting Information.

To improve the photophysical properties, deepen the energy levels, and improve sublimation behavior, we developed two further new Eu(II)-based materials utilizing aza-18-crown-6 ethers as ligands, which have not been reported to date. In **EuCrown**, divalent europium is coordinated by three $OR_2$ and three $HNR_2$ donors of a symmetric 4,10,16-triaza-18-crown-6 ether, which allows two $[CB_{11}H_{12}]^-$ carborate anions to electrically neutralize divalent europium from above and below the plane of the crown ether in a pseudo-$C_{3h}$ symmetric structure. Subsequently, the material was obtained by mixing the Eu(II)-carborate with commercially available ligands in an appropriate solvent in a ratio of 1:1 under inert conditions, as detailed in the Supporting Information, Section S1. While the resulting symmetric structure (*cf.* Figure 2) helps to reduce the dipole moment and improve volatility, the (HSAB-) soft and weakly coordinating carborate anions ensure deep energy levels and thus help to stabilize Eu(II) against oxidation. The intended improvements are confirmed by DFT, which predicts a low dipole of 2.7 D, along with considerably higher IE and ES–IE of 5.8 eV and 2.8 eV, respectively. With this ES–IE, ambipolar mCP-based hosts and electron-transporting hosts with triazine or pyridine groups with LUMO levels around −2.7 eV appear, at least in theory, compatible. The predicted emission energy of 2.8 eV is still higher than the experimental value of 2.7 eV (in TAPC, *cf.* Section 2.2), but the deviation is much smaller than for EuCrypt.

To further improve the design, we aimed to attach the carborate anions covalently to a coordinating ligand. Therefore, we developed **EuCovCrown**, where two $[CB_{11}H_{12}]^-$ anions are linked from the boron vertex opposing the carbon atoms to the nitrogen atoms of a 7,16-diaza-18-crown-6 ether via a saturated B-OCC-N bridge (*cf.* Figure 2). Accordingly, the synthesis of the desired ligand requires selective manipulation on the 12-boron-vertex of a $[CB_{11}H_{12}]^-$ anion. Although many protocols have been reported on the substitution of the closo-1-carborate anion,[20,21] most of them describe manipulations at the 1-carbon-vertex. The selective substitution on the 12-position is more challenging and often suffers from selectivity issues or a limited scope of possible substitutions. However, selective oxidation of the anion to introduce a hydroxy group at the desired position is well-known and investigated,[22,23] which enables functionalization by simple alkylation chemistry. Thus, the desired ligand was accessible by nucleophilic substitution using the 12-oxido-carborate and the chloro-acetylated derivative of the 18-crown-6 ligand, followed by amide reduction.

This synthetic route enables the preparation of EuCovCrown, in which the carborate anions are covalently attached to the crown-ether ligand, further improving the emitter design. EuCovCrown was prepared by protonolysis of the ammonium salt of the ligand, using EuHMDS as the lanthanide source (further details are provided in the Supporting Information, Section S1). Since the resulting structure is less symmetric, the predicted dipole is slightly larger with 4.9 D, which is countered by the rigidifying cyclohexyl bridges that are expected to promote volatility. The IE and ES–IE are slightly shallower than those of EuCrown, with values of 5.5 eV and 2.6 eV, which is presumably due to the additional partially negatively charged oxygen atoms coordinating Eu. However, in exchange for the shallower levels, the covalently attached anions and higher coordination of Eu improve the chemical stability of the complex and shield the Eu core against nucleophilic attack (*vide infra*). EuCovCrown can be seen as a compromise in that some of the deep energy levels provided by the carborate anions are exchanged for higher stability and better shielding. Figure 2 summarizes the molecular structures and calculated energy levels of the Eu(II) emitters studied in this work.

Both EuCrown and EuCovCrown exhibit highly efficient deep-blue emission in the solid state as well as in dispersions. Photoluminescence studies in dilute dispersions in a toluene:THF solvent mixture reveal single-band blue emission with peak emission at 450-460 nm, as expected for Eu(II) 4*f*–5*d* transitions. EuCrown shows a high PLQY of 90% and an excited-state lifetime of 820 ns, while EuCovCrown exhibits a PLQY of 88% with a longer lifetime of 950 ns. The combination of single-band blue emission, long excited-state lifetime, and high photoluminescence quantum yield confirms the formation of Eu(II) emitter complexes and the highly efficient nature of the 4*f*–5*d* emission of these emitters. However, photoluminescence properties obtained in dilute dispersions of the emitters are not directly comparable to those in thin-film or OLED environments. Further details for the PLQY measurements are provided in Section S3 of the Supporting Information.

In addition to emission characteristics, sufficient thermal stability and sublimation behavior are essential requirements for emitter materials intended for vacuum deposition, which is the industrial standard for OLED production. Thermogravimetric analysis (TGA) and dedicated sublimation experiments confirm that both EuCrown and EuCovCrown exhibit sublimation yields of approximately 78%. Both vacuum-train-sublimed emitters provide stable deposition rates under ultra-high vacuum similar to established OLED functional layers, rendering them compatible with industrial OLED manufacturing. Detailed thermal stability and sublimation results, as well as data of fabricated OLEDs incorporating these emitters are provided in Sections S2 and S4 of the Supporting Information.

## 2.2 Photoluminescence of Eu(II) Emitters in Different Host Environments

To enable a precise analysis of Eu(II) emission behavior in different solid-state environments while minimizing system complexity and confounding factors, we investigated the photoluminescence characteristics of EuCrypt, EuCrown, and EuCovCrown in thin films of representative OLED host materials. Thin films with emitter concentrations of 5, 20, and 40 wt% were prepared by thermal evaporation under ultra-high vacuum using TAPC, SiDBFCz, and B3PyPB as hosts, which span a broad range of electron-accepting LUMO energies (−2.0, −2.5, and −2.8 eV, respectively) as well as distinct coordinating characters. Steady-state and time-resolved photoluminescence measurements were employed to monitor changes in intrinsic emission characteristics, specifically emission spectra and excited-state lifetimes, which serve as sensitive probes of host-dependent radiative and non-radiative processes of the emitter. Details on host selection, energy-level alignment, and spectroscopic characterization are provided in Section S5 of the Supporting Information.

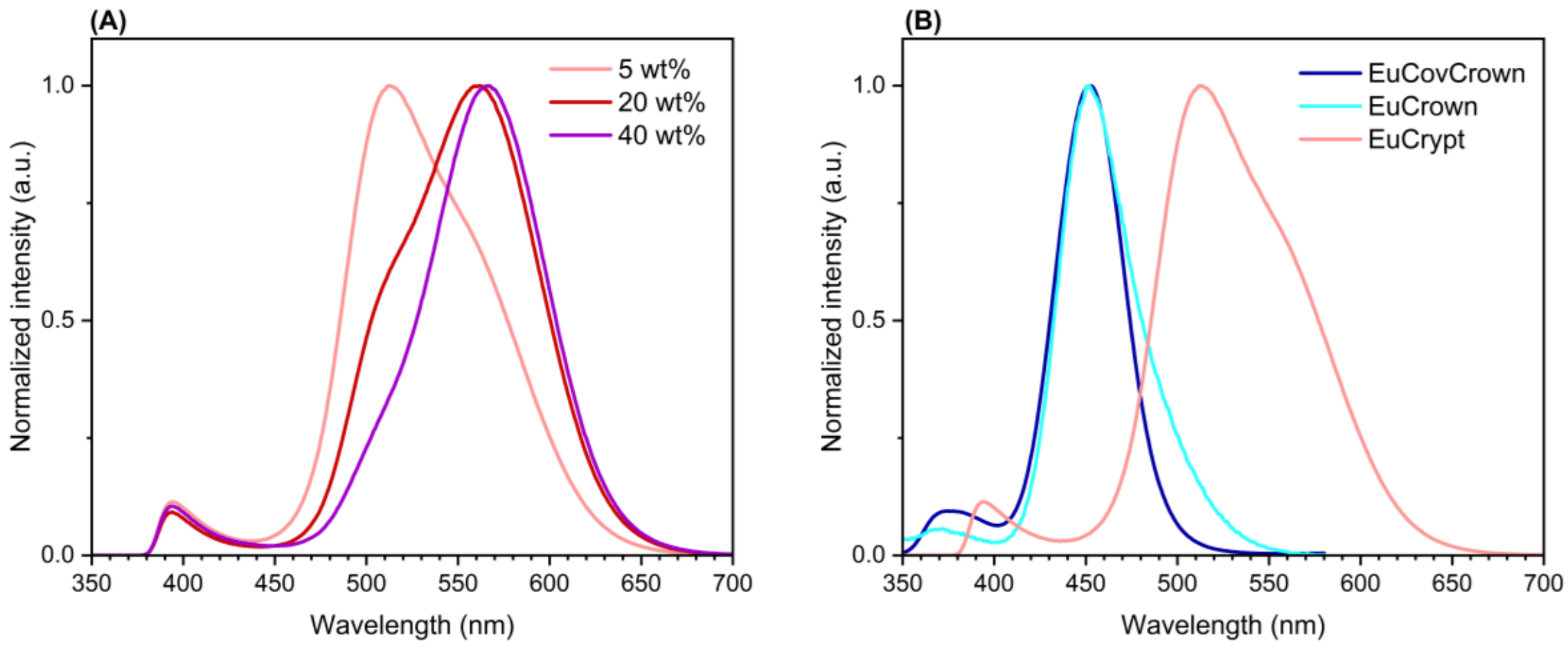

**Figure 3.** Photoluminescence spectra of Eu(II) emitters in TAPC host matrix. **(A)** Concentration-dependent photoluminescence spectra of EuCrypt in TAPC. **(B)** Comparison of the emission spectra of EuCrown and EuCovCrown with EuCrypt in TAPC at a concentration of 5 wt%. The steady-state photoluminescence spectra were obtained from thin films excited at a wavelength of 300 nm, which excites both host and emitter. The weak, high-energy emission observed below ~400 nm originates from residual host emission.

We first examine the emission properties of the reference emitter EuCrypt. **Figure 3**A shows the emission of EuCrypt at various concentrations in TAPC. EuCrypt exhibits a broad emission with two emission bands peaking at approximately 515 nm and 560 nm, corresponding to green and yellow, respectively. As shown in Figure 3A, the intensity ratio of these peaks varies with the concentration of the emitter in the film. At the lower concentration of 5 wt%, the green emission is the highest peak with a pronounced yellow shoulder. With increasing concentration to 20 wt%, the yellow emission becomes prominent and dominates the spectrum at 40 wt% (*cf.* **Table 1**). The excited-state lifetimes for EuCrypt at both recorded emission peaks at various concentrations in TAPC are provided in Table 1. Each concentration shows different lifetime values for both green and yellow components, with the yellow component always being longer than the green component. This suggests that there is an energy transfer mechanism coupling the two excited states. A detailed analysis of the excited-state dynamics of EuCrypt is outside the scope of this manuscript.

EuCrypt shows similar concentration-dependent emission in SiDBFCz host (see Supporting Information, Section S5). In B3PyPB, the emitter shows no detectable emission. The recorded emission peaks for the green and yellow components corroborate the findings reported in ref. [12], where Liu et al. showed the emission color to vary with the molecular environment and morphology in the solid state. The peaks observed for EuCrypt are 515 nm and 560 nm, corresponding to 2.4 eV and 2.2 eV, respectively. Since these energies are well below the lowest triplet energies of the hosts, exciton energy transfer can be excluded as a quenching mechanism (see Supporting Information, Section S5).

**Table 1**. Photophysical properties of EuCrypt at different concentrations in the TAPC host.

| Emitter concentration | 5 wt% | 20 wt% | 40 wt% |
|---|---|---|---|
| $\lambda_{max}$ [nm] | 512 | 562 | 567 |
| FWHM [nm] | 90 | 97 | 75 |
| $\tau_{green}$ [ns] | 362 | 177 | 101 |
| $\tau_{yellow}$ [ns] | 503 | 507 | 423 |

$\lambda_{max}$: peak emission wavelength; FWHM: full width at half maximum; $\tau_{green}$: excited state lifetime for the green-peak emission (at 514 nm); $\tau_{yellow}$: excited state lifetime for the yellow-peak emission (at 560 nm).

The emission characteristics of EuCrown and EuCovCrown are largely concentration independent compared to EuCrypt, with the peak emission wavelength remaining unchanged with minor variations in the excited-state lifetime (see Supporting Information, Section S5). The lack of a pronounced concentration dependency and a narrower emission in the blue region indicates that the EuCrown and EuCovCrown are less susceptible to their molecular environment than EuCrypt. Therefore, further comparison of the Eu-based emitters in different host materials was carried out at a fixed concentration of 5 wt%, chosen to simplify the analysis and avoid unnecessary repetition across host systems.

While EuCovCrown features a single-Gaussian emission spectrum peaking at 452 nm, EuCrown shows a slightly broader emission at longer wavelengths (see Gaussian Fitting in Supporting Information, Section S5). Notably, the emission of EuCrown is strongly suppressed in B3PyPB. Comparing the emission of EuCrown and EuCovCrown in TAPC shown in Figure 3B reveals that EuCrown (FWHM of 48 nm) exhibits a broader emission profile than EuCovCrown (FWHM of 43 nm). The narrower emission of EuCovCrown could be related to a decrease in the freedom of motion of the anions due to being covalently attached to the ligand of the complex.

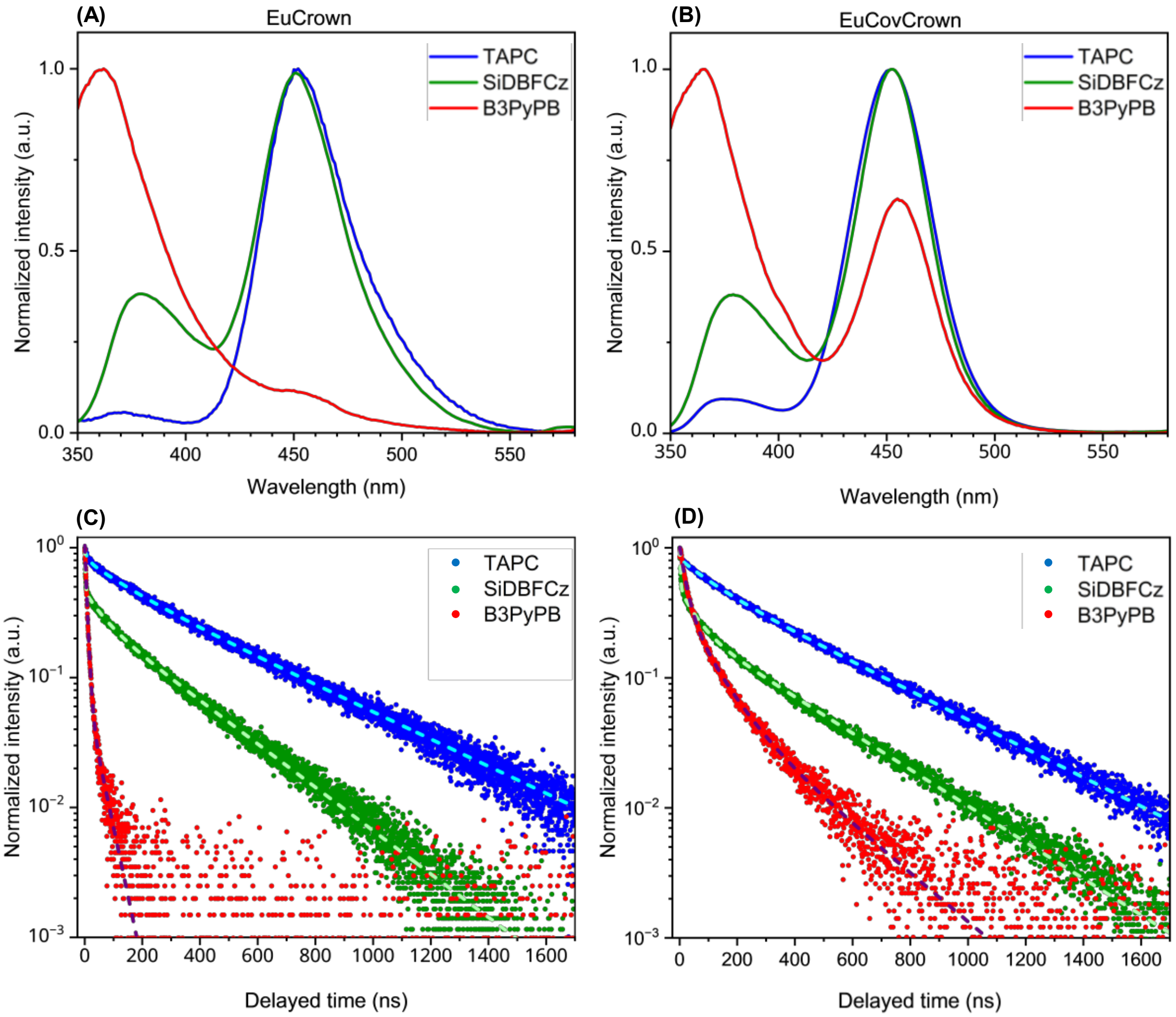

**Figure 4.** Photoluminescence characteristics of Eu(II) emitters in different host matrices. Steady-state photoluminescence spectra **(A, B)** and excited-state decay curves **(C, D)** of EuCrown (left) and EuCovCrown (right) at 5 wt% concentration in TAPC, SiDBFCz, and B3PyPB hosts. The steady-state photoluminescence spectra were obtained from thin films excited at 300 nm. Excited-state decay curves were measured using time-correlated single-photon counting (TCSPC), with a 294 nm pulsed laser used for excitation. The monochromator of the detector was set to the peak emission wavelength of each emitter, as determined from the steady-state spectra (or to 450 nm for EuCrown:B3PyPB), for signal detection and data acquisition. The emission at ~450 nm is assigned to Eu-based emission, whereas the high-energy emission originates from the host.

To investigate the influence of the host matrix on the emission of Eu-based emitters, steady-state emission spectra and excited-state decay curves of EuCrown and EuCovCrown are shown in **Figure 4**. The corresponding peak emission wavelengths and excited-state lifetimes extracted from these measurements are summarized in **Table 2**. Figure 4A shows the emission spectra of EuCrown in the three hosts. Each spectrum consists of two major contributions; the higher-energy one can be attributed to host emission. This remaining host emission indicates that energy transfer from host to emitter is not complete, which is partially attributed to the low oscillator strength of the Eu(II) complexes and further to unfavorable spectral overlap between donor host emission and acceptor emitter absorption.[24–26] The TAPC:EuCrown emission spectrum (blue curve in Figure 4A) shows a barely visible host contribution, and the emitter contribution is clearly visible. The SiDBFCz:EuCrown spectrum, represented by the green curve, shows both host and emitter contributions, while the B3PyPB:EuCrown spectrum, represented by the red curve, consists mostly of host emission, with a small side peak at around 450 nm.

**Table 2.** Spectral and decay properties of EuCrown and EuCovCrown in different host matrices.

| Emitter | Host | $\lambda_{max}$ [nm] | FWHM [nm] | $\tau_1$ [ns] ($B_1$) | $\tau_2$ [ns] ($B_2$) | $\tau_3$ [ns] ($B_3$) |
|---|---|---|---|---|---|---|
| | TAPC | 452 | 48 | 394 (0.69) | 68 (0.19) | — |
| EuCrown | SiDBFCz | 450 | 48 | 289 (0.26) | 94 (0.18) | 4 (0.52) |
| | B3PyPB | — | — | 103 (0.02) | 9 (0.98) | — |
| | TAPC | 453 | 43 | 388 (0.64) | 88 (0.21) | — |
| EuCovCrown | SiDBFCz | 452 | 39 | 331 (0.24) | 70 (0.23) | 4 (0.50) |
| | B3PyPB | 455 | 40 | 237 (0.09) | 86 (0.29) | 21 (0.66) |

$\lambda_{max}$: peak emission wavelength of the Eu complex (host emission excluded); FWHM: full width at half maximum; $\tau_n$: excited-state lifetimes of the emitter peak emission ($\lambda_{max}$, or 450 nm for EuCrown:B3PyPB); and $B_n$ denotes the relative amplitude of each lifetime component.

Similar to EuCrown, the emission spectra of EuCovCrown in the three hosts are plotted in Figure 4B. Analogous to EuCrown, EuCovCrown shows emission at 452 nm in TAPC and SiDBFCz. In all three samples, the contribution of the host is visible and appears identical to the EuCrown case. However, unlike EuCrown, EuCovCrown shows a clear emitter contribution in B3PyPB, although the host emission remains dominant.

One possible way to explain the behavior of the emitters in B3PyPB is to consider their energy levels with respect to electron-confinement. Specifically, B3PyPB has a LUMO of −2.8 eV, while EuCrown and EuCovCrown have ES–IEs of 2.8 eV and 2.6 eV, respectively. Since the LUMO of the host is very close to (EuCrown) or deeper (EuCovCrown) than the binding energy of the excited electron (ES–IEs) of the emitters, combined with the extended nature of the diffuse 5*d*-orbitals, an electron transfer from emitter to host may occur, resulting in the formation of a charge-transfer (CT) excited state between host and emitter creating a competitive non-radiative decay channel. Such a case would be analogous to CT-state formation between two materials, provided there is a favorable energy level arrangement. Because the LUMO energy of B3PyPB and ES–IE of the emitter are equal for EuCrown, it should be less affected by quenching stemming from host-emitter interaction based on a new state formation, compared to EuCovCrown, where the host LUMO is deeper by 0.2 eV. With this in mind, both TAPC and SiDBFCz feature LUMO energies of −2.0 eV and −2.5 eV, respectively, which are smaller than the ES–IEs for both EuCrown and EuCovCrown. This effectively prevents the CT-state formation, rendering the CT-based quenching pathway ineffective.

To clarify the behavior of the new Eu-based emitters in different host environments, excited-state decay measurements were performed, as shown in Figure 4C and 4D. The extracted data from the decay measurements are summarized in Table 2. In Figure 4C, the TAPC:EuCrown decay curve (blue) shows a bi-exponential decay behavior, with lifetime values of 68 ns and 394 ns (*cf.* Table 2). The longer, dominant 394 ns component in the decay is attributed to the emission from EuCrown. The shorter 68 ns component is not fully understood to date. It either relates to a minor fraction of EuCrown emission that is influenced by quenching or non-linear effects, or it originates from host triplet states influenced by the Eu complex via the heavy-atom effect (see Section S5 of the Supporting Information for host triplet emission spectra). However, analyzing the continuous wave spectra in Figure 4A, no additional spectral features can be observed. The SiDBFCz:EuCrown decay curve (green) shows a tri-exponential decay with components of 4 ns, 94 ns, and 289 ns (*cf.* Table **2**). Again, the long-lived 289 ns component corresponds to the intrinsic emission from EuCrown. The fast 4 ns component is assigned to prompt host fluorescence, which overlaps with the host emission and hence leaks into the detection window. The mechanism for the intermediate 94 ns component is not clear, but may be connected to a fraction of EuCrown sites decaying faster due to non-linear processes or quenching, again with no additional spectral features suggesting a distinct emission band, in particular in view of the comparable weights of both components (0.26 and 0.18, *cf.* Table 2). Equally, the B3PyPB:EuCrown decay curve (red) shows predominantly fast host fluorescence with a lifetime of 9 ns (dominant with a 0.98 weight). The additional 103 ns component is observable but very weak (0.02, *cf.* Table 2). Noticeably, no longer-lived component that could be assigned to EuCrown is detected in this host.

The decay behavior of EuCovCrown, shown in Figure 4D, is closely analogous to that of EuCrown in TAPC and SiDBFCz hosts, with a similar assignment of the contributions. In TAPC, a bi-exponential decay is observed with components of 88 ns and 388 ns, respectively, with the latter being the dominant one (0.64 of weight). In SiDBFCz, the decay is tri-exponential (4 ns, 70 ns, and 331 ns), where the fastest component originates from direct host fluorescence and the longest is connected to the intrinsic EuCovCrown emission.

The major difference between EuCrown and EuCovCrown in these two hosts is that EuCovCrown (331 ns) shows a longer decay component than EuCrown (289 ns) in SiDBFCz, while both emitters show nearly identical decay lifetimes in TAPC (394 ns and 388 ns for EuCrown and EuCovCrown, respectively). As also seen in the emission spectra, the emitters are different in their behavior in B3PyPB as well. Here, EuCovCrown displays a tri-exponential decay, with lifetimes of 21 ns, 86 ns, and 237 ns. Similar to the other hosts, the components are assigned to direct host emission, intermediate emission that cannot be assigned without doubt, and intrinsic emitter emission, respectively. This can be understood as the emitter being affected less strongly by B3PyPB compared to EuCrown, where no emitter component was observed.

Considering the photoluminescence decays, it seems unlikely that a CT-state formation is the only reason for the loss or reduction of emission for EuCrown and EuCovCrown in SiDBFCz and B3PyPB. The reasons are the following: Firstly, the decay curves clearly show that both emitters are affected by SiDBFCz, leading to reduced luminescence and reduced PLQY (see Section S5 of the Supporting Information for further discussion). Taking the decay components in TAPC as a reference for minimally perturbed emission, SiDBFCz is estimated to reduce the PLQY of EuCrown and EuCovCrown by 27 and 15 percentage points, respectively. Secondly, EuCovCrown has a lower ES–IE in the same range as the LUMO of B3PyPB, which should make it more susceptible to CT-state formation. However, the opposite trend is observed: EuCrown is quenched more strongly in B3PyPB than EuCovCrown. Therefore, another effect must be at play that relates to a different aspect of these Eu(II) complexes.

### 2.3 DFT Investigation of Eu(II) Emitter-Host Interactions

To explore this effect, we first built a minimal model of the B3PyPB host, reducing it to the subunit meta-Phenyl-Pyridine (mPhPy), and conducted conformer searches for all three emitters with this additional neutral molecule attached to the outside of the complex. The employed CREST workflow (see Computational Details) relies on metadynamic simulations at the GFN2-xTB level to explore and sample the conformational space and uses a spherical wall potential to avoid a complete separation.[27,28] This creates a large ensemble with hundreds of conformers within an energy cutoff of 12 kcal mol$^{-1}$ (0.52 eV) that includes structures with the pyridine moiety of mPhPy coordinating europium instead of or in addition to the anions (this is required to allow topology changes at the Eu center, see Section S6 of the Supporting Information), as well as structures with mPhPy non-covalently attached to the outside of the complex. Using CENSO, this initial ensemble is filtered with DFT of increasing sophistication and eventually a robust r$^2$SCAN-3c optimization.[28]

Application of this protocol showed a major difference between EuCrown on the one hand, and EuCrypt and EuCovCrown on the other: Only for EuCrown, the CREST/CENSO workflow finds an energetically favorable structure where Eu(II) is coordinated by pyridine in addition to the carborate anion (ΔG = 1.4 kcal mol$^{-1}$ or −0.06 eV) compared to the outside-adduct, *i.e.*, with mPhPy not coordinating Eu. Inspection reveals that for EuCrypt and EuCovCrown, mPhPy-coordinated structures are considered in the initial steps (CREST) but filtered out during the refinement (CENSO), meaning that they are energetically unfavorable (ΔG≫0). From this result, we conclude that in EuCrypt and EuCovCrown, the metal center is better shielded by the ligands (and anions) than in EuCrown, where mPhPy insertion is favorable.

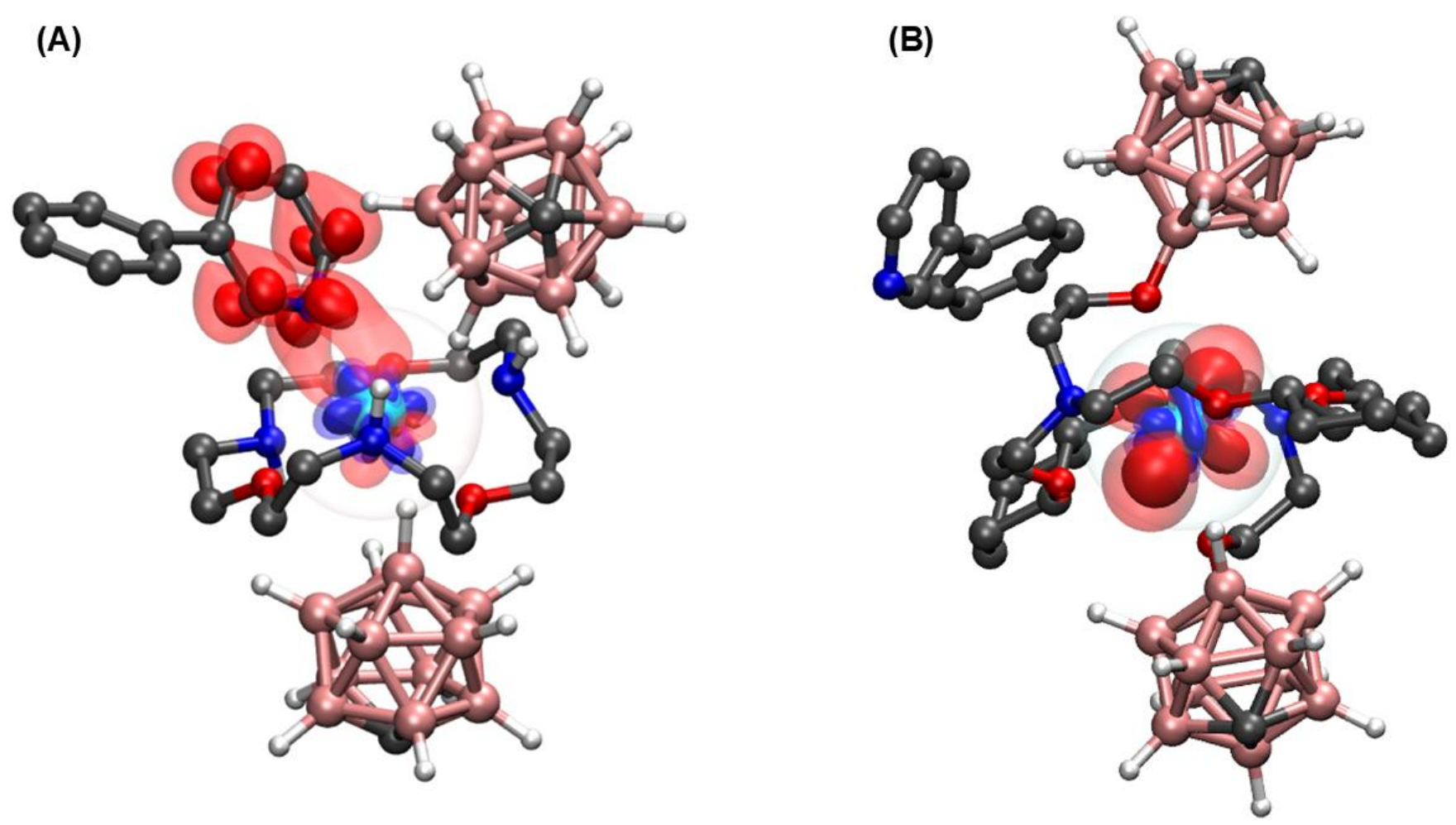

**Figure 5.** Difference-density plots of the lowest excited states of mPhPy–Eu(II) emitter adducts. **(A)** mPhPy–EuCrown adduct and **(B)** mPhPy–EuCovCrown adduct, with mPhPy located on the upper-left side and the Eu(II) emitter on the right side of each structure. Regions of decreasing electron density are shown in blue, and regions of increasing electron density are shown in red. Visualizations were generated using VMD 1.9.2.[19]

Exploring the photophysical properties of the EuCrown species coordinated by mPhPy explains the lack of emission in this host despite matching energy levels: the ground-state structure with the Eu-coordinating mPhPy exhibits a substantial shift in the absorption energy from 3.2 eV to 2.9 eV, which is further enhanced in optimizing the lowest excited state, yielding an emission energy of merely 1.81 eV (*cf.* 2.86 eV for the isolated EuCrown). Moreover, during the excited-state optimization, the Eu-N(pyridine) distance decreases substantially from 2.68 Å (ground state equilibrium) to 2.32 Å. Inspection of the difference densities of the lowest excited state at the excited-state (shown in **Figure 5**A) and ground state structure reveals the reason for this drastic change: While at the ground-state geometry of the mPhPy-coordinated EuCrown the lowest excited state is dominated by the $d-f$ transition with a minor metal-to-ligand CT (MLCT) admixture, optimization of this state yields an almost pure MLCT state in which an electron from Eu is transferred to the LUMO of mPhPy (with a weak $d-f$ admixture). This is in stark contrast to the mPhPy adducts of EuCrypt and EuCovCrown: As evident from panel B of Figure 5, the mPhPy remains outside the coordination sphere in EuCovCrown, where it only slightly alters the energy levels (largest change ±0.15 eV), but does not change the nature of the lowest excited $4f{-}5d$ state. Apparently, in EuCrown, the coordination by the mPhPy model for B3PyPB drastically alters the properties of the emitters, favoring irreversible oxidation to Eu(III) via excited-state electron transfer to the coordinating pyridine. This manifests in the emergence of a low-energy MLCT excited state, which corresponds to a loss of electron confinement in the excited state. This hypothesis is further supported by previous reports about a quenching effect of pyridine ligands for Ln emitters.[29]

Having identified the characteristics of the MLCT state in the mPhPy-coordinated EuCrown, we conducted further TD-DFT calculations for the mPhPy-outside (non-coordinating) adducts of all three emitters with a larger number of excited states (100). These calculations locate the higher-lying MLCT states at energies of 5.8 eV for EuCrown, 5.3 eV for EuCovCrown, and 4.8 eV for EuCrypt, which is in line with the ordering of the ES–IE of the emitters. Note that in these calculations, the energy of the MLCT state is severely overestimated relative to the $d-f$ excited state because of several, mostly technical reasons. These include an incomplete description of CT states by the employed LR-PCM solvent model [30,31], the employed minimal host model, and the lack of any geometric relaxation in the excited state. Nevertheless, the large shift of the energy of the MLCT state between the outside- and Eu-coordinating mPhPy adduct in EuCrown clearly illustrates that loss of electron confinement not only depends on the static energy levels (host LUMO and emitter ES–IE), but also on the ability of the emitter and its ligand to keep nucleophilic host molecules away from the Eu center. This may be described as steric shielding of the metal center against its environment.

To corroborate these results and directly demonstrate the superior shielding of EuCovCrown (and EuCrypt), **Figure 6** summarizes the results of another numerical experiment which employs iodide anion as “steric probe” to model an approaching host molecule: This is motivated by the fact that iodide has no internal degrees of freedom and is attracted to Eu(II), which enables very simple 1-D Eu-iodide distance scans. The potential energy surfaces (PESs) obtained in these scans are displayed in Figure 6A. To obtain them, we placed the iodide at a distance of 7 Å from Eu, moving it towards Eu(II) in steps of 0.2 Å while all other degrees of freedom are optimized/relaxed at each step.

Starting with the EuCrypt complex (orange line), two coordinated iodide ligands are present, while the position of a third approaching iodide is indicated by a red circle. Inspection reveals a global minimum at around 5 Å (0.75 eV below the point at infinite distance, middle right structure), which corresponds to iodide binding to the NH protons not coordinated by the other two iodide atoms. Further decreasing the Eu-I distance increases the energy by about 0.1 eV (2 $\mathrm{kcal\,mol^{-1}}$) as hydrogen bonds (not shown) are broken, and the other two iodides have to be pushed away from Eu, yielding the local minimum at around 4 Å (middle left structure).

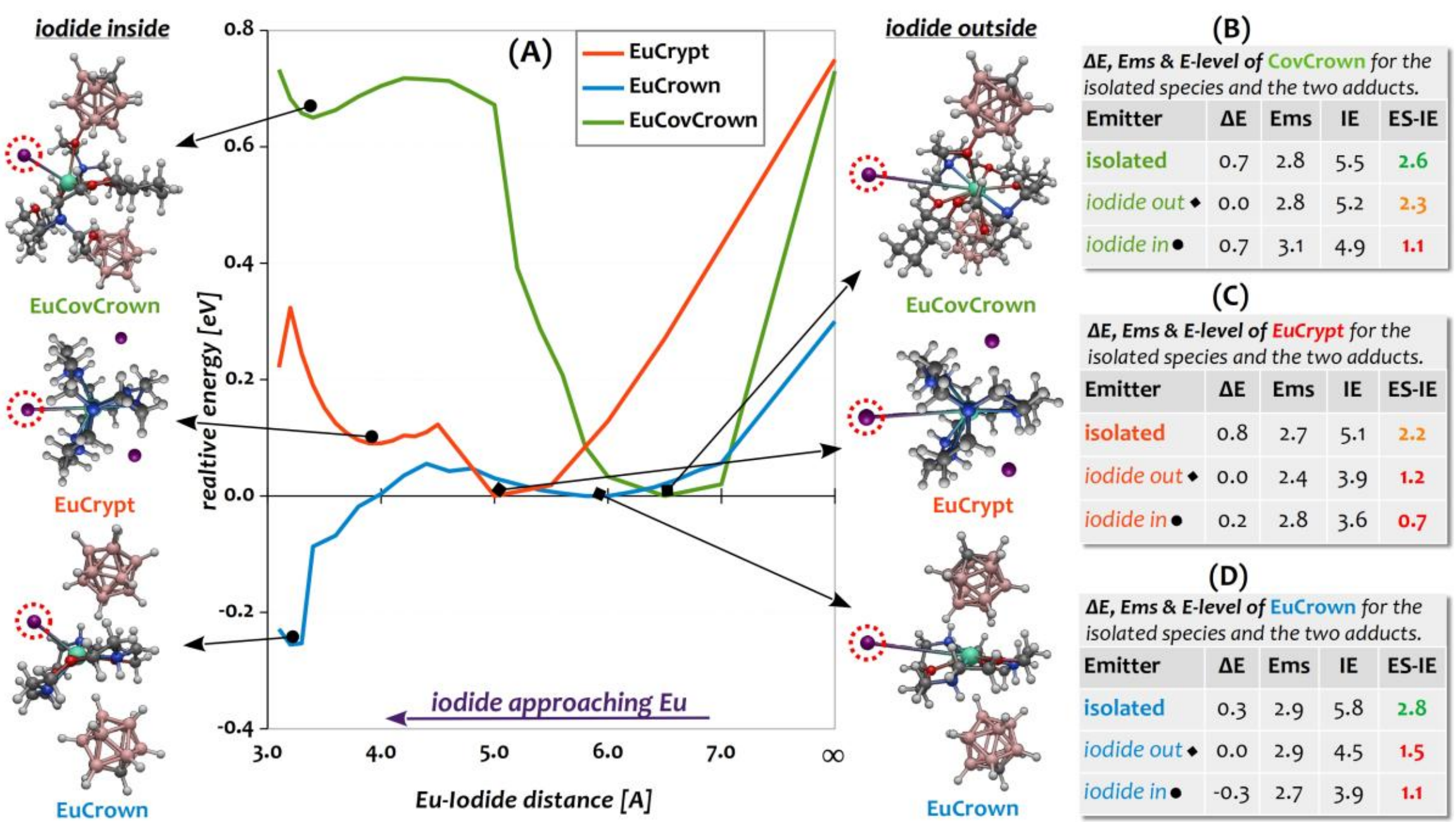

ΔE, Ems & E-level of CovCrown for the isolated species and the two adducts.

| Emitter | ΔE | Ems | IE | ES-IE |
| --- | --- | --- | --- | --- |
| isolated | 0.7 | 2.8 | 5.5 | 2.6 |
| iodide out ◆ | 0.0 | 2.8 | 5.2 | 2.3 |
| iodide in ● | 0.7 | 3.1 | 4.9 | 1.1 |

ΔE, Ems & E-level of EuCrypt for the isolated species and the two adducts.

| Emitter | ΔE | Ems | IE | ES-IE |
| --- | --- | --- | --- | --- |
| isolated | 0.8 | 2.7 | 5.1 | 2.2 |
| iodide out ◆ | 0.0 | 2.4 | 3.9 | 1.2 |
| iodide in ● | 0.2 | 2.8 | 3.6 | 0.7 |

ΔE, Ems & E-level of EuCrown for the isolated species and the two adducts.

| Emitter | ΔE | Ems | IE | ES-IE |
| --- | --- | --- | --- | --- |
| isolated | 0.3 | 2.9 | 5.8 | 2.8 |
| iodide out ◆ | 0.0 | 2.9 | 4.5 | 1.5 |
| iodide in ● | -0.3 | 2.7 | 3.9 | 1.1 |

**Figure 6**. Steric shielding of Eu(II) emitters against iodide coordination. **(A)** Relaxed scan of the potential energy surface (PES) of iodide approaching Eu(II) in EuCrypt (orange), EuCrown (blue), and EuCovCrown (green) at the DFT/r$^2$SCAN-3c/SMD(toluene) level of theory. The energetically most favorable outside-adduct of all emitters is set to zero. Structures corresponding to the local minima identified as iodide-outside adduct (marked with a diamond, right) and iodide-inside adduct (marked with a circle, left) are shown. **(B-D)** Show the energy difference $\Delta E$ (relative to the iodide-outside adduct), emission energy (Ems), IE, and ES–IE for the isolated species, iodide-outside, and iodide-inside adducts to illustrate the impact a charge in the proximity of the emitter has on the energy levels. ES–IE values are color-coded where green indicates good electron confinement, orange borderline confinement, and red certain loss of confinement.

In contrast, EuCrown (blue line) shows the most favorable interaction with iodide, as evident from the deep global minimum corresponding to the iodide-inside adduct (bottom left structure with an Eu-iodide distance of only 3.2-3.3 Å). Moreover, there is only a small barrier < 0.1 eV that separates the broad local minimum at ≈6 Å (iodide-outside, bottom right structure) from the global minimum characterized through the direct coordination of Eu(II). Apparently, the crown-ether ligand poorly shields the central cation. This is consistent with the differences observed for the interaction with the host model mPhPy, which exhibits a very similar structure to the iodide-adduct (*cf.* Figure 5).

Finally, in EuCovCrown (green line), the ligand effectively prevents the addition of iodide, creating a large barrier (0.7 eV or 16 kcal mol$^{-1}$) and a highly endothermic reaction energy (0.6 eV or 13 kcal mol$^{-1}$) for converting the iodide-outside adduct to the iodide-inside adduct (top left structure with an Eu–I distance of ≈3.4 Å). The mechanistic reason for this improved shielding is that it is not possible to simply add iodide to the ligand-sphere (like in EuCrown), but because of the more bulky and rigid ligand, the addition of iodide forces one of the carborate anions to detach from Eu (*cf.* top left structure in Figure 6). Again, this is consistent with the results for the mPhPy model, for which we did not find any low-energy adducts in which pyridine is coordinating the central Eu.

To illustrate the impact of the coordination on the energy levels, we have composed the inlays B-D in Figure 6, which provide energy difference (relative to the outside adduct), emission energy, and energy levels calculated for the isolated emitter and the two adducts. Inspection of the evolution of the ES–IE values of the three emitters shows a large decrease of $>1$ eV for the iodide-outside structure for EuCrypt and EuCrown, which contrasts with the smaller decrease of only 0.3 eV for EuCovCrown. These results demonstrate that while the excited-state ionization energy (ES–IE) defines the necessary energetic boundary for electron confinement, emission quenching in B3PyPB is ultimately governed by coordination-induced loss of confinement.

Overall, since the triplet energies of the host materials exceed the emission energies of the Eu(II) emitters, triplet exciton quenching is unlikely to be the deactivation pathway. The observed quenching in the host-environment thin films is plausibly associated with charge-transfer-state formation through electron transfer between host and emitter, facilitated by the spatially extended character of the Eu(II)-centered 5*d* excited state. Such a process requires favorable energy alignment and sufficient orbital overlap for charge transfer, which may arise from coordination interactions between nucleophilic host motifs and the Eu(II) center and is influenced by the degree of steric shielding around the Eu(II) center. In this framework, the ES–IE defines a necessary energetic boundary for electron confinement, while coordination-induced orbital overlap provides a possible pathway for loss of confinement and non-radiative deactivation. In this context, the reduced sensitivity of EuCovCrown to host-induced quenching is consistent with more effective steric shielding of the Eu core, which limits detrimental coordination interactions.

However, despite the improved shielding of EuCovCrown, its reduced lifetime in B3PyPB films indicates that further optimization of the ES–IE is needed to fully exploit the potential of these emitters in OLED host environments. This highlights a key challenge for the further development of Eu-based emitters and devices: deep electronic energy levels (requiring weakly interacting anions) must be combined with rigid steric shielding (requiring Eu to be coordinatively saturated) in a molecule that retains the deep and pure blue color emission and volatility of EuCrown and EuCovCrown. Such a deep and well-shielded Eu(II) complex would be more tolerant toward the selection of hosts, enabling efficient blue-emitting devices.

## 3 Summary and Conclusions

In this work, we report two novel Eu(II) emitters, EuCrown and EuCovCrown, based on crown-ether ligands and carborate anions, combining single-band deep-blue 4*f*–5*d* photoluminescence (PLQY ≈ 90%) with sufficient thermal stability and sublimation yields to enable reliable vacuum deposition, a prerequisite that has been particularly challenging for deep-blue Eu(II) emitters. To demonstrate processability and device compatibility, representative OLEDs were fabricated (see Section S4 of the Supporting Information), confirming the functionality of these emitters in electroluminescent devices. Both emitters exhibit deep-blue electroluminescence ($\lambda_{max} \leq 458$ nm), with the best-performing device achieving a maximum external quantum efficiency exceeding 12%. These results demonstrate that efficient and color-pure deep-blue electroluminescence can be realized from Eu(II)-based emitters while maintaining the processing requirements necessary for OLED integration.

Beyond emitter development, we provide a combined experimental and theoretical analysis of the factors governing Eu(II) emitter performance in OLEDs, highlighting fundamental differences from conventional organic emitters. We establish a multilevel computational workflow (SQM/DFT/TD-DFT) that yields ES–IE and related properties in good agreement with experiments. The analysis provides a mechanistic picture of how host molecules in the emissive layer interact with the europium center, and how such interactions impair Eu(II) emission characteristics. In particular, coordination of nucleophilic host motifs can induce low-lying charge-transfer states, compromise electron confinement, and thereby quench Eu(II) luminescence. Suppressing these interactions requires both effective steric shielding of the Eu center and sufficient energetic confinement of the excited 5*d* electron, highlighting key design principles for maintaining Eu(II) emission efficiency.

Taken together, these insights demonstrate the feasibility of deep-blue luminescence from vacuum-processable Eu(II) complexes and establish a framework for the rational design of next-generation Eu(II) emitters that enables the translation of the favorable parity-allowed 4*f*–5*d* atomic transitions of divalent europium into higher-efficiency vacuum-processed blue OLEDs.

## 4 Experimental and Computational Protocols

### Experimental Setup and Measurements

Organic materials and starting materials for synthesis were obtained from commercial suppliers and used as received unless otherwise noted. Thin films and OLED devices were fabricated on clean quartz substrates or pre-patterned ITO-coated glass substrates using a Kurt J. Lesker ultra-high-vacuum thermal evaporation system. Device encapsulation was carried out inside a nitrogen-filled glovebox using a glass cavity cap, UV-curable epoxy, and a getter. Complete layer structures and fabrication parameters are given in the Supporting Information, Section S4.

Photoluminescence and PLQY measurements of emitters were performed under an inert atmosphere using an Edinburgh Instruments spectrometer and a Hamamatsu Quantaurus-QY instrument. Host–emitter thin-film PL measurements were carried out using a Horiba Fluoromax-4 spectrofluorometer inside a nitrogen-filled glovebox. Steady-state PL spectra were recorded with excitation wavelengths of 300 nm, while excited-state lifetimes were measured using time-correlated single-photon counting (TCSPC) with 294 nm pulsed excitation. The emitted photons were detected using a monochromator–photomultiplier tube setup, with the detection wavelength set to the peak emission wavelength of the Eu(II) emitter and a detection bandwidth of approximately 5-10 nm, allowing for selective detection of emitter-related emission. Additional experimental details and data analysis procedures are provided in the Supporting Information section S5.

**Computational Workflow and Details**

Throughout this study, we used the ORCA [32] program package version 6.0.1 for ground- and excited-state geometry optimizations with DFT and TD(A)-DFT, and the CREST program version 2.12 [27,28,33] and CENSO version 1.2.0 for conformer sampling.[28]

Conformational searches were conducted for all emitters using GFN2-xTB [34] metadynamic simulations and optimizations as implemented in CREST [27,28,33] with a slightly modified GFN2 Hamiltonian and CREST parameters (NCI mode, alpb [35] solvation with parameters for toluene, increased energy cut-off of 12 $\mathrm{kcal\,mol^{-1}}$). The resulting conformer ensembles were prescreened (CENSO part0) at the PBE-D4/def2-SV(P)+gCP level of theory with an energy-cutoff of 10 $\mathrm{kcal\,mol^{-1}}$,[36–39] and eventually screened (CENSO part1) and reranked with $r^2$SCAN-3c/SMD(toluene) with a cut-off of 7 $\mathrm{kcal\,mol^{-1}}$,[40,41] followed by optimizations (CENSO part 2) of the lowest 16 conformers with $r^2$SCAN-3c level of theory with an energy cut-off of 3 $\mathrm{kcal\,mol^{-1}}$. The lowest conformer resulting from this protocol was used in all further steps.

Absorption (abs), emission (ems), ionization (IE), and excited-state ionization (ES–IE) energies are calculated for the lowest conformers with spin-unrestricted time-dependent (TD) density functional theory (DFT) in the Tamm-Dancoff approximation [42,43] using the range-separated hybrid functional $\omega$B97X-D3 with default settings,[44] the SMD solvent model [41] for toluene combined with the LR-PCM approach for non-equilibrium solvation as implemented in ORCA6,[32] the def2-TZVP/ECP basis set and on Eu, the def2-SVP/ECP basis set on all other atoms,[45,46] and the geometric counter-poise (gCP) correction of Grimme.[38] The motivation to use $\omega$B97X-D3 is threefold: Firstly, it accurately predicts emission energies of Eu(II) complexes (many more were tested than shown here in this work). Secondly, it correctly recovers the 1/r asymptote of charge transfer states, which are a central topic in this work. Thirdly, it predicts the HOMO/ionization energy (IE) of the employed hosts in very good agreement with UPS experiments (see Supporting Information, Sections S3 and S5). Further details on the protocols of the calculations (e.g., inclusion of nuclear relaxation, treatment of relativistic effects) can be found in the Supporting Information, Section S6.

**Conflict of Interest**

M.S., P.I., J.-M.M., F.K., and C.R. are employed at beeOLED GmbH, which works towards commercialization of lanthanide materials for OLED applications. K.L. and S.R. support beeOLED GmbH as scientific advisors. The mentioned authors hold options in beeOLED GmbH.

**Data Availability Statement**

The data that support the findings of this study are available from the corresponding author upon reasonable request.

## Supporting Information

# High-Efficiency Deep Blue Single-Gaussian Europium(II) Emitters and their Emitter-Host Interactions

*Mahmoud Soleimani,* [a,b] *Paulius Imbrasas,* [b] *Jan-Michael Mewes,* [b] *Felix Kaden,* [b] *Stephanie Anna Buchholtz,* [a] *Karl Leo,* [a] *Sebastian Schellhammer,* [a] *Carsten Rothe,* [b] *Sebastian Reineke* [a]

[a] Institute of Applied Physics (IAP) and Dresden Integrated Center for Applied Physics and Photonic Materials (IAPP), Technische Universität Dresden, Dresden, Germany
[b] beeOLED GmbH, Dresden, Germany

## S1 Preparation of Emitters

### S1.1 General Information

All reactions were performed under an inert nitrogen atmosphere in flame-dried glassware using standard Schlenk-techniques or a glovebox. tetrahydrofuran (THF, 99.5%, water < 50 ppm), diethyl ether ($Et_2O$, 99.5%, water < 50 ppm), Acetonitrile (MeCN, 99.9%), and toluene (99.85%, water < 50 ppm) were purchased from Fisher Scientific. THF was purified by passage through a column of molecular sieves followed by aluminum oxide (basic, Brockmann-I). Triethylamine ($NEt_3$) was freshly distilled over calcium hydride ($CaH_2$). Trimethylammonium *closo*-1-carbadodecaborate (>97%) was purchased from Katchem and used without further purification. 1,7,13-Trioxa-4,10,16-triazacyclooctadecane (3NH, 96%) and 2,8,15,21-Tetraoxa-5,18-diazatricyclo[20.4.0.0$^{9,14}$] hexacosane ($2NH_2Ch$) were purchased from Chemieliva and purified by distillation prior to use. All other commercially available reagents were used as received, unless otherwise stated. Europium bis(bis(trimethylsilyl)amide)($Eu(HMDS)_2$) and Trimethylammonium 12Hydroxy1carbadodecaborate were prepared according to the literature. NMR spectra were recorded on a Magritek Spinsolve 60. Chemical shifts ($\delta$) are reported in parts per million (ppm) downfield of tetramethylsilane. The complexes were characterized by mass spectrometry (MS) in solution (methanol), recorded on a Thermo Scientific MSQ Plus Mass Detector; elemental analysis (EA), performed on an Elementar UNICUBE; thermogravimetric analysis (TGA), performed on a NETZSCH STA 449 F5 Jupiter, and their photophysical properties.

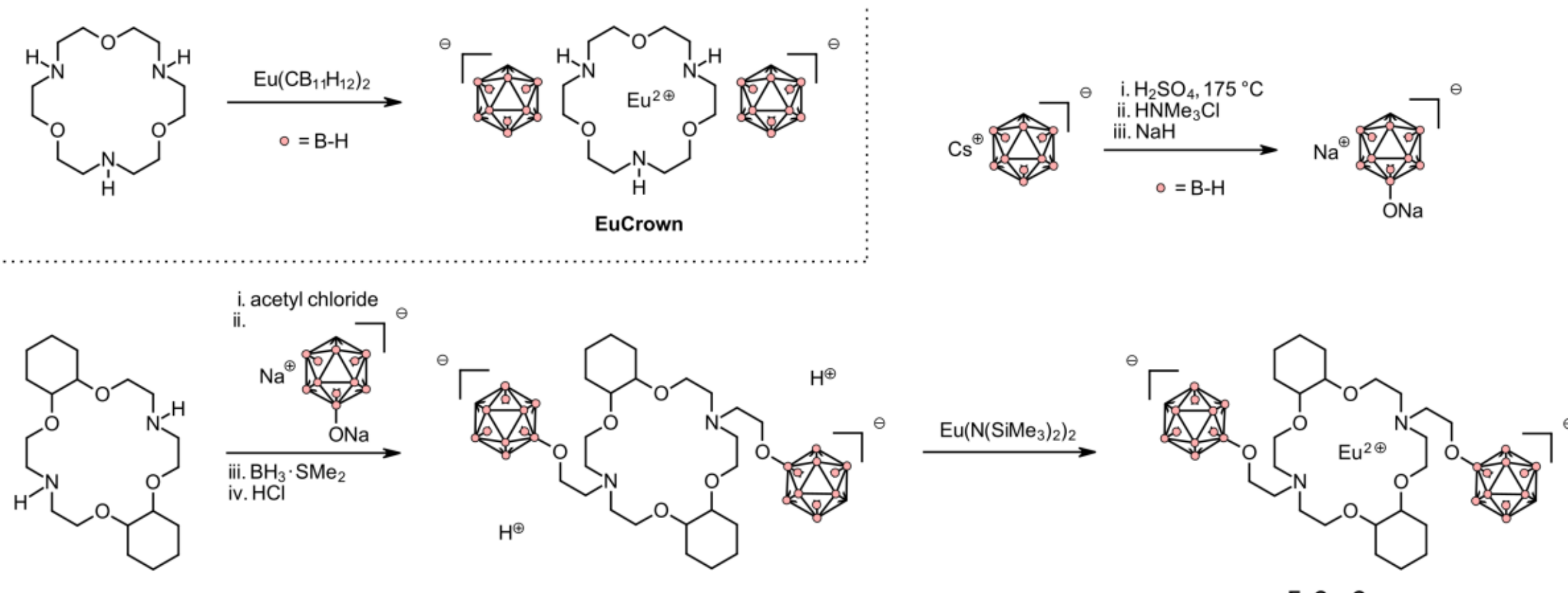

**Figure S1.** Schematic representation of the synthesis of EuCrown and EuCovCrown. $Eu(HMDS)_2$ first reacts with trimethylammonium closo-1-carbadodecaborate to form $Eu(CB_{11}H_{12})_2$, which upon coordination with either 1,7,13-trioxa-4,10,16-triazacyclooctadecane or a functionalized diazatrioxa-macrocycle yields EuCrown and EuCovCrown, respectively.

## S1.2 Preparation of EuCrown

1. Europium bis(closo-1-carbadodecaborate) tetrahydrofuran ($Eu(CB_{11}H_{12})_2$)

A solution of 1.17 g (5.76 mmol, 2.0 eq.) trimethylammonium closo-1-carbadodecaborate in THF (20 mL) was added dropwise to a solution of 1.78 g (2.88 mmol, 1.0 eq.) $Eu(HMDS)_2$ in THF (40 mL). The resulting suspension was stirred for 1 h at room temperature and then filtered. The resulting powder was washed with THF and dried under vacuum to yield the title compound as a white powder in 97%. This product was used without further purification.

2. Europium bis(closo-1-carbadodecaborate) 1,7,13-trioxa-4,10,16-triazacyclooctadecane (EuCrown)

A solution of 484 mg (1.85 mmol, 1.0 eq.) 3NH in toluene (20 mL) was added dropwise to a solution of 945 mg (1.85 mmol, 1.0 eq.) $Eu(CB_{11}H_{12})_2$ in toluene (120 mL). The resulting suspension was stirred for 1 h at room temperature. The solvent was removed in vacuum, and the resulting powder was washed with $Et_2O$ and dried under vacuum to give EuCrown in 77% yield. For further purification, the obtained powder was sublimed at 320 °C ($1.7 \times 10^{-6}$ mbar). Final yield: 78%. **MS (ESI):** m/z = 284.4 ($[3NH + Na]^+$); 262.4 ($[3NH + H]^+$); 239.1 ($[M - 2\ (CB11H12)^+\ 2\ MeOH]^{2+}$); 223.2 ($[M - 2\ (CB11H12)^+\ MeOH]^{2+}$). **Elemental analysis:** Calculated: C, 12.10; H, 4.32; N, 6.00. Found: C, 11.76; H, 5.06; N, 6.02.

## S1.3 Preparation of EuCovCrown

1. Preparation of sodium 12-oxido-1-carbadodecaborate (NaO-$CB_{11}H_{11}$)

A solution of 1.36 g (6.22 mmol, 2.2 eq.) of trimethylammonium 12-hydroxy-1-carbadodecaborate in THF (25 mL) was treated with 0.75 g (18.7 mmol, 6.6 eq.) sodium hydride (60% in mineral oil). The mixture was stirred at room temperature for 90 min. The precipitate was filtered off, washed with THF, and the filtrate concentrated to dryness under vacuum. The off-white powder was redissolved in THF (25 mL), and the solution was used directly in whole for step 2b.

2. Preparation of the ligand for EuCovCrown

a) 2.0 g (5.4 mmol, 1.0 eq.) $2NH_2Ch$ was dissolved in MeCN (60 mL) and treated with 1.50 mL $NEt_3$ (10.80 mmol, 2.0 eq.). The mixture was cooled to 0 °C, and 0.86 mL (10.80 mmol, 2.0 eq.) of chloroacetyl chloride was added dropwise. The ice bath was removed, and the reaction mixture was stirred overnight at room temperature. The precipitate was filtered off, and the filtrate concentrated to dryness. The crude product was purified by flash chromatography (EtOAc, aluminum oxide, neutral Brockmann-I) to yield 2 g (72%) of a colorless oil.

b) 1.48 g (2.83 mmol, 1.0 eq.) of this colorless oil was dissolved in THF (20 mL). The previously prepared solution of NaO-$CB_{11}H_{11}$ in THF (step 1) was added via syringe. The mixture was stirred at room temperature overnight. The reaction was quenched by the addition of an aqueous solution of NaCl (half-sat., 20 mL). Stirring continued for 30 min. The layers were separated, the aqueous layer extracted with EtOAc (4 × 20 mL), and the combined organic layers concentrated to dryness under vacuum. The residue was dissolved in a boiling mixture of $H_2O$/EtOH (4:3, 140 mL) and treated with trimethylamine hydrochloride (5 eq, 1351 mg, 14.1 mmol). The mixture was cooled in an ice bath for 1 h. The white precipitate was filtered off, washed with $H_2O$, and dried under vacuum at 50 °C. 2 g (80%) of a white powder was yielded.

c) The white solid (2.0 g, 2.25 mmol, 1.0 eq.) was dissolved in THF (45 mL). The mixture was stirred at room temperature, and 11.25 mL (2 M, 22.5 mmol, 10 eq.) of a borane-dimethylsulfide complex solution in THF was added via syringe. The mixture was heated to reflux for 24 h. The reaction was cooled to room temperature, and stirring continued at room temperature for 40 h. The reaction was quenched carefully with hydrochloric acid (4 M, 20 mL). The mixture was heated to reflux for 30 min. The solution was concentrated to dryness. The residue was re-diluted with $H_2O$ (125 mL), heated to reflux and filtered hot. The remaining solid was dried in a vacuum and purified by flash chromatography on silica gel ($CH_2Cl_2$/MeCN, 20-100%). The obtained crude product was washed with MeOH and dried under vacuum to give 1.12 g (43%) of the ligand as its hydro salt.

**$^1$H{$^{11}$B} NMR** (60 MHz, acetone-d$_6$): $\delta$[ppm] = 8.91 (br. s, 2H), 3.99-3.69 (m, 8H), 3.69-3.42 (m, 10H), 3.30-2.90 (m, 10H), 2.33-1.85 (m, 8H), 1.78–1.29 (m, 22H), 1.27–0.91 (m, 8H). **$^{11}$B{$^1$H} NMR** (20 MHz, DMSO): $\delta$[ppm] = 12.05 (br. s, 2B), –15.81 (br. s, 20B). **$^{13}$C{$^1$H} NMR** (16 MHz, DMSO): $\delta$[ppm] = 81.14, 63.55, 61.99, 53.58, 50.35, 29.51, 23.42. **MS (ESI):** = 369.9 ($[M-2H]^{2-}$); 741.4 ($[M-H]^{-}$); 763.5 ($[M-2H+Na]^{-}$). **Elemental analysis:** Calculated: C, 44.64; H, 10.07; N, 6.51. Found: C, 44.86; H, 10.02; N, 6.74.

3. Preparation of EuCovCrown

To a solution of 958 mg (1.29 mmol, 1.0 eq.) of the ligand in THF (20 mL) was added a solution of 795 mg (1.29 mmol, 1.0 eq.) $Eu(HMDS)_2$ in THF (8 mL). The mixture was stirred at room temperature for 3 h. The mixture was concentrated under vacuum to ca. half of its volume, the precipitate was filtered, washed with small amounts of THF and toluene, and dried under vacuum. EuCovCrown was isolated as a colorless powder (610 mg, 89% yield). Further purification was achieved by sublimation at 290 °C under high vacuum ($2.5 \times 10^{-6}$ mbar). Final yield: 77%.

**MS (ESI):** m/z = 369.9 ($[M-Eu]^{2-}$). **Elemental analysis:** Calculated: C, 34.98; H, 7.45; N, 3.14. Found: C, 35.10; H, 7.63; N, 3.10.

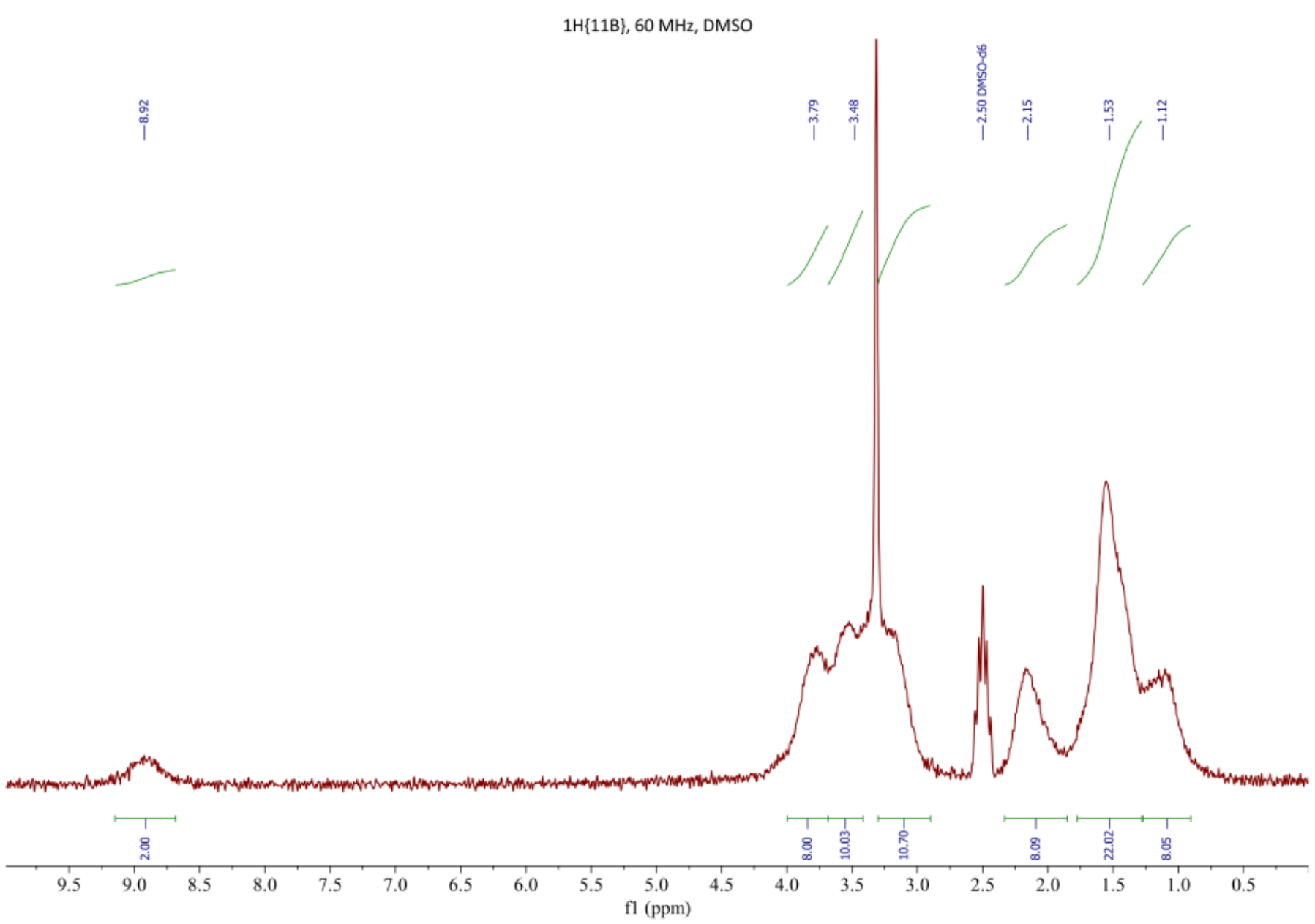

**Figure S2.** $^{1}H\{^{11}B\}$ NMR spectrum of the ligand for EuCovCrown.

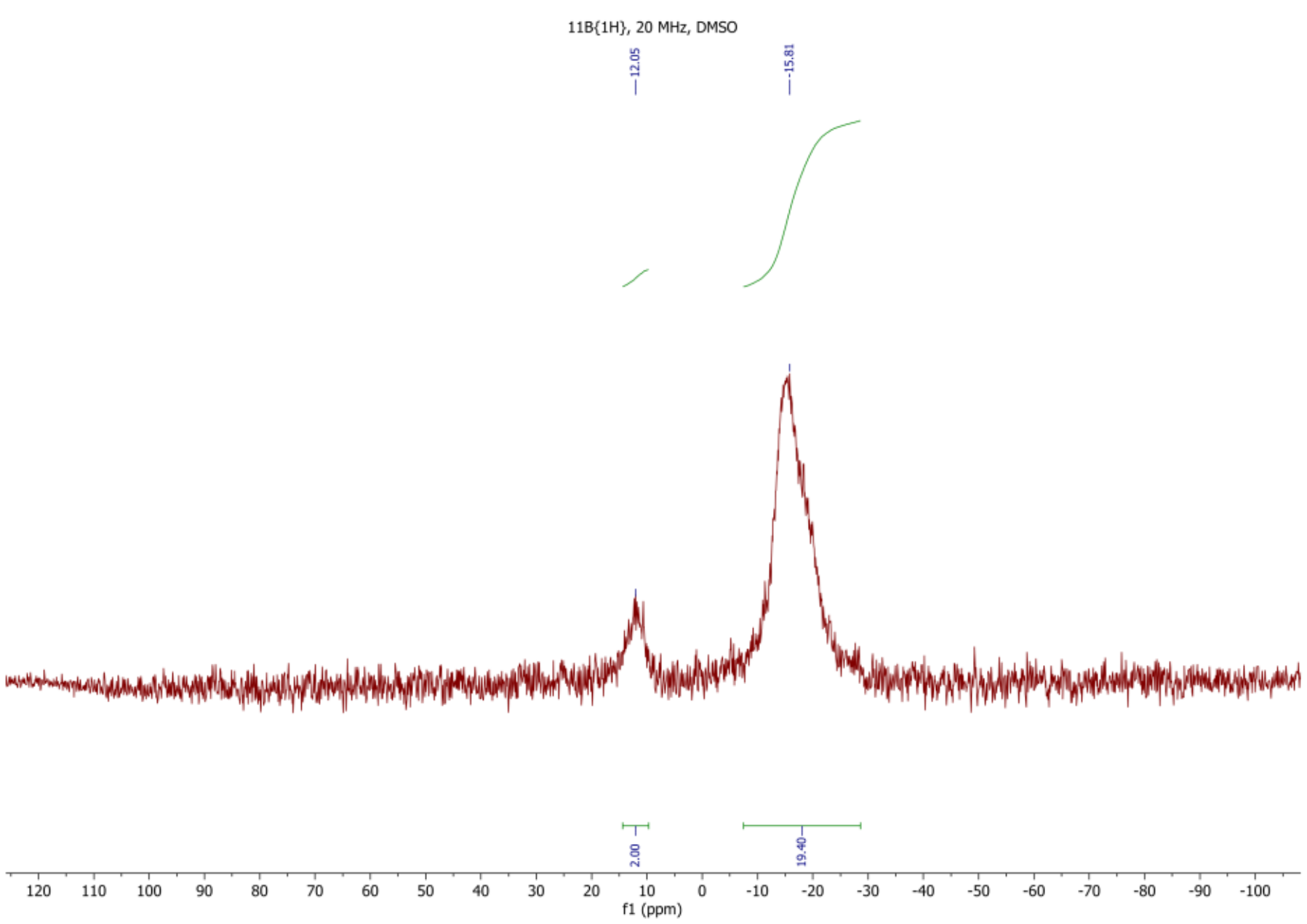

**Figure S3.** $^{11}B\{^{1}H\}$ NMR spectrum of the ligand for EuCovCrown.

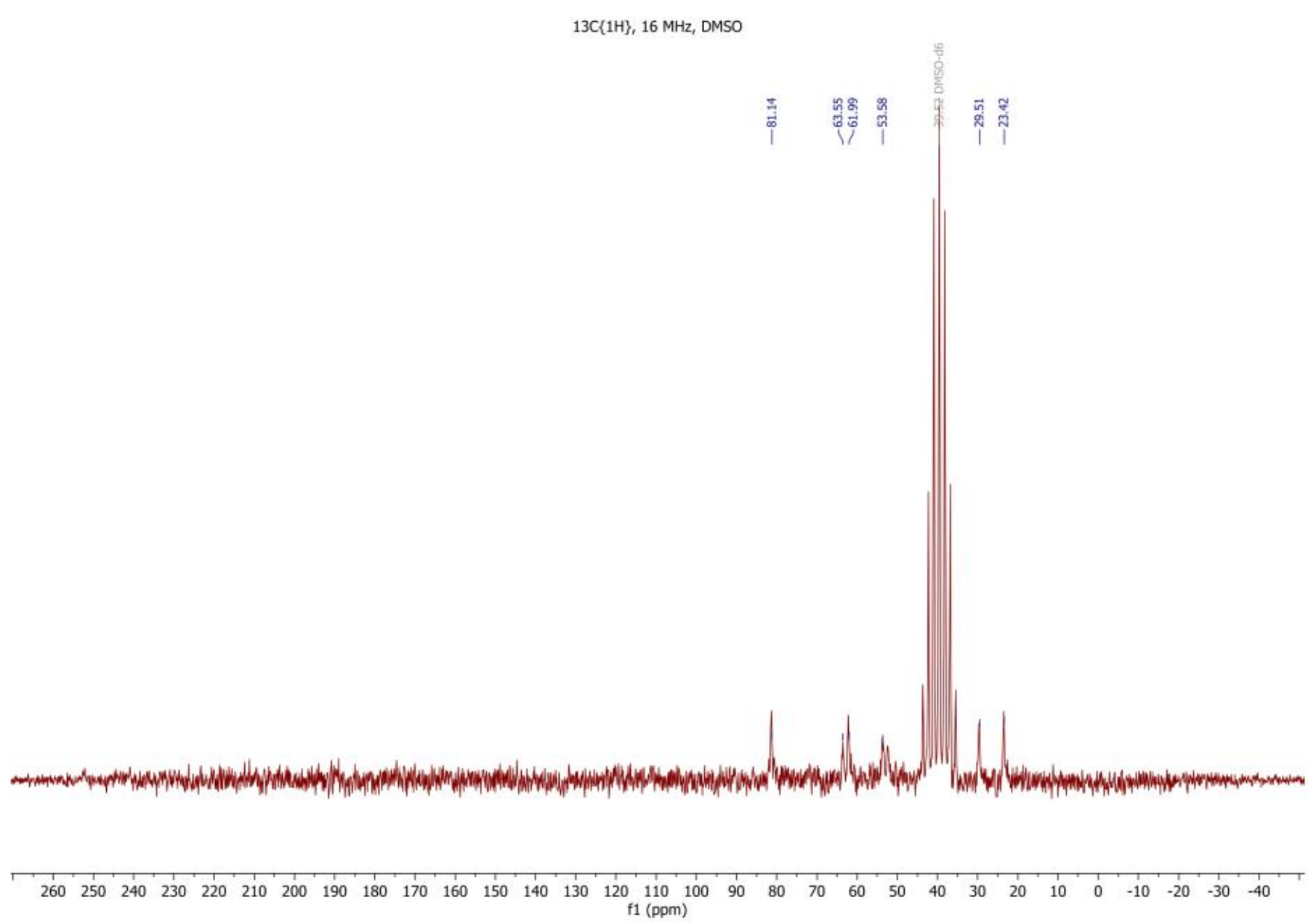

**Figure S4.** $^{13}C\{^{1}H\}$ NMR spectrum of the ligand for EuCovCrown.

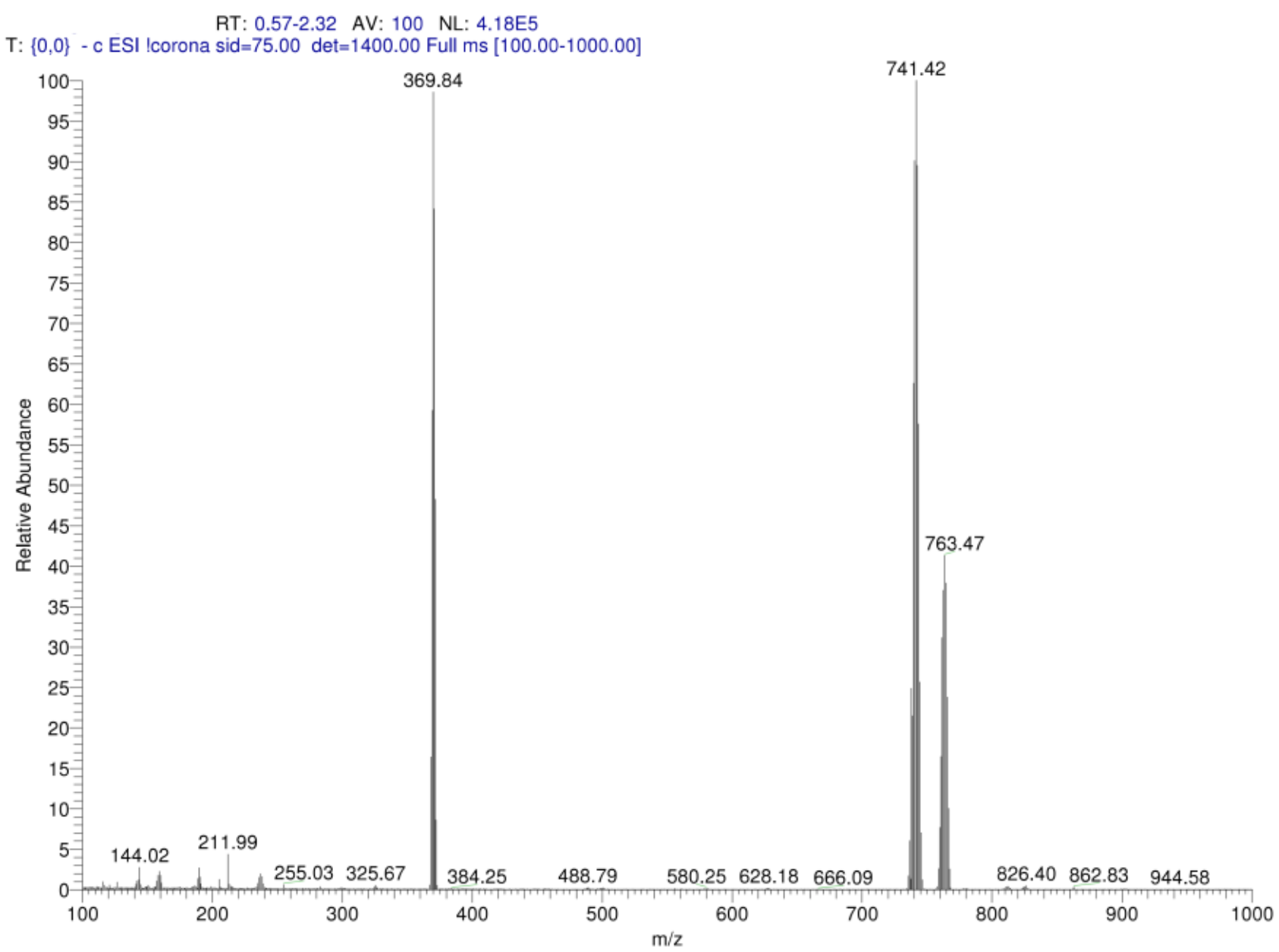

**Figure S5**. ESI-MS spectrum of the ligand for EuCovCrown.

## S2 Thermal Properties and Sublimation

The thermal stability of EuCrown and EuCovCrown was investigated by thermogravimetric analysis (TGA, Figure S6). Measurements were performed using 40 µL aluminum crucibles without lids under vacuum with a heating rate of 10 K $min^{-1}$ and with typical sample masses of 2–10 mg. The temperature corresponding to 5% mass loss, $T_{5\%}$, was determined to be approximately 349 °C for EuCrown and 311 °C for EuCovCrown. Upon further heating to temperatures up to 450 °C, pronounced mass loss was observed, which is attributed to evaporation accompanied by partial decomposition. Complete mass loss was not achieved in either case, suggesting the presence of residual non-volatile components or decomposition products formed at elevated temperatures.

To directly assess the suitability of both materials for vacuum processing, controlled sublimation experiments were performed. The compounds were loaded into sublimation tubes under an inert atmosphere and processed using an mBraun DSU05 sublimation system. The system was evacuated to pressures in the range of $10^{-5}$–$10^{-6}$ mbar and heated stepwise. The optimal sublimation temperatures were identified as $T_{sub}$ = 320 °C for EuCrown and $T_{sub}$ = 290 °C for EuCovCrown, corresponding to mass losses of 2.0% and 3.8%, respectively, according to the TGA results. At temperatures below $T_{sub}$, no changes in material appearance or optical properties were observed. In contrast, heating more than 10 °C above $T_{sub}$ resulted in an increase in pressure, discoloration, and loss of luminescence after prolonged heating, indicating the onset of thermal decomposition during sublimation. Under optimized conditions, both materials could be sublimed with a sublimation yield of approximately 78% for both complexes.

Under OLED fabrication conditions, the evaporation tool operates at pressures several orders of magnitude lower than those used in the dedicated sublimation experiments. As a result, both EuCrown and EuCovCrown begin to evaporate at significantly lower temperatures, typically in the range of 200–220 °C, depending on the chamber pressure. Following the onset of evaporation, the deposition rate can be adjusted and maintained over a wide operating range suitable for OLED fabrication, while remaining well below the thermal decomposition thresholds determined by TGA, thereby providing a wide operating margin for vacuum deposition.

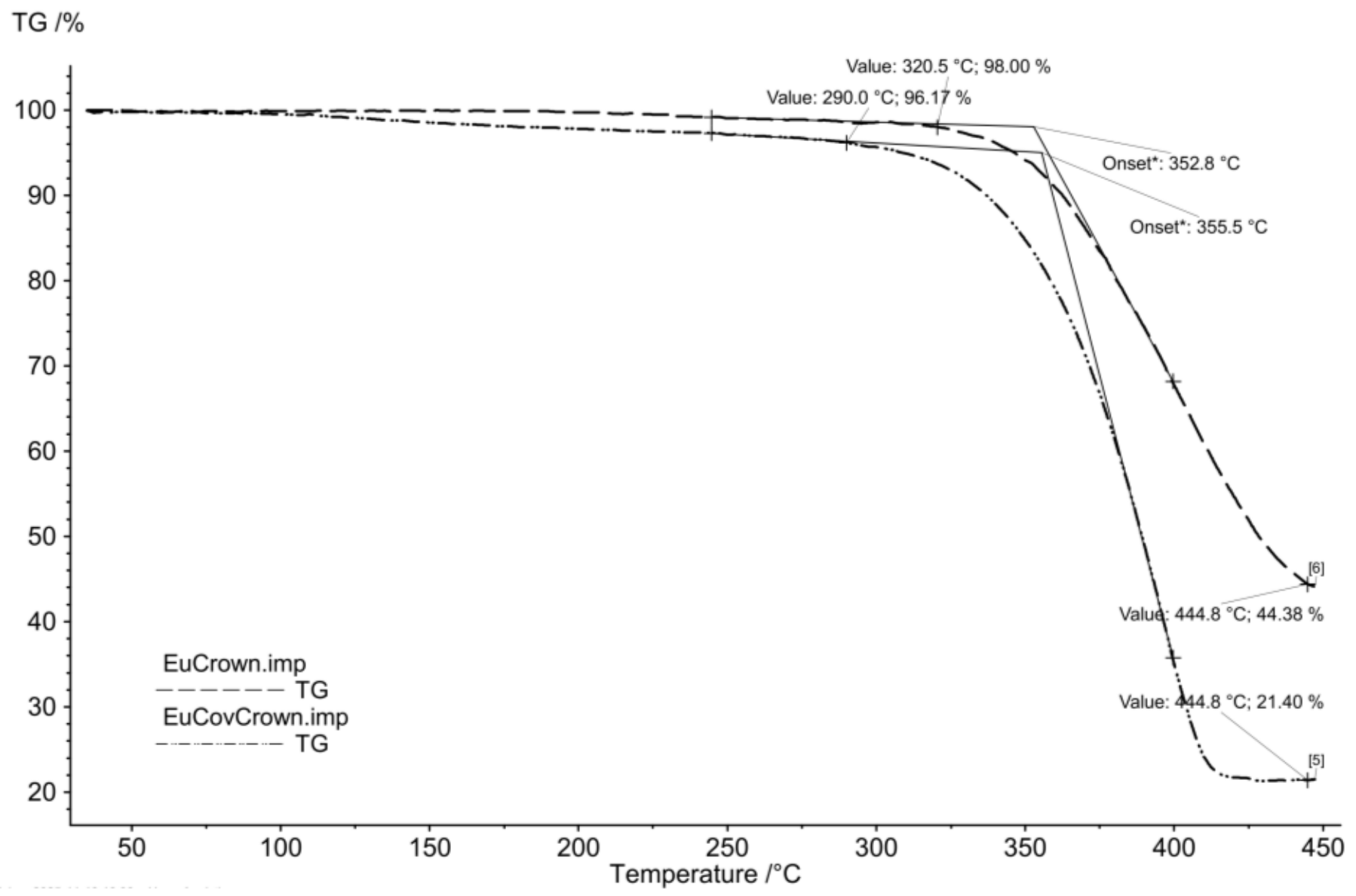

**Figure S6.** TGA data of EuCrown (dashed line) and EuCovCrown (dash-dotted line).

## S3 Characterization of Eu(II) Emitters

Photoluminescence quantum yield (PLQY) measurements were performed in dilute dispersions prepared in a mixed toluene/tetrahydrofuran (THF) solvent system (3:1 v/v). Due to the very low solubility of the Eu(II) complexes, the measurements were conducted using well-dispersed solid particles rather than solutions, with the amount of emitter corresponding to an effective concentration of approximately $10^{-5}$ M. The emission spectra of the emitters in low-concentration dispersions are shown in **Figure S7**. Absolute PLQY values of 90% for EuCrown and 88% for EuCovCrown were obtained using a Hamamatsu integrating-sphere–based photoluminescence quantum yield measurement system (Quantaurus-QY). Time-resolved photoluminescence measurements yielded excited-state lifetimes of 820 ns for EuCrown and 950 ns for EuCovCrown and were performed using an Edinburgh Instruments spectrometer with an excitation wavelength of 310 nm. The corresponding radiative and non-radiative decay rate constants derived from the PLQY and lifetime data are summarized in **Table S1**. Although the measured values in dispersions reflect the radiative efficiency of the Eu(II) cores, they are used only as an initial reference for emitter luminescence and are not indicative of their performance in thin-film environments.

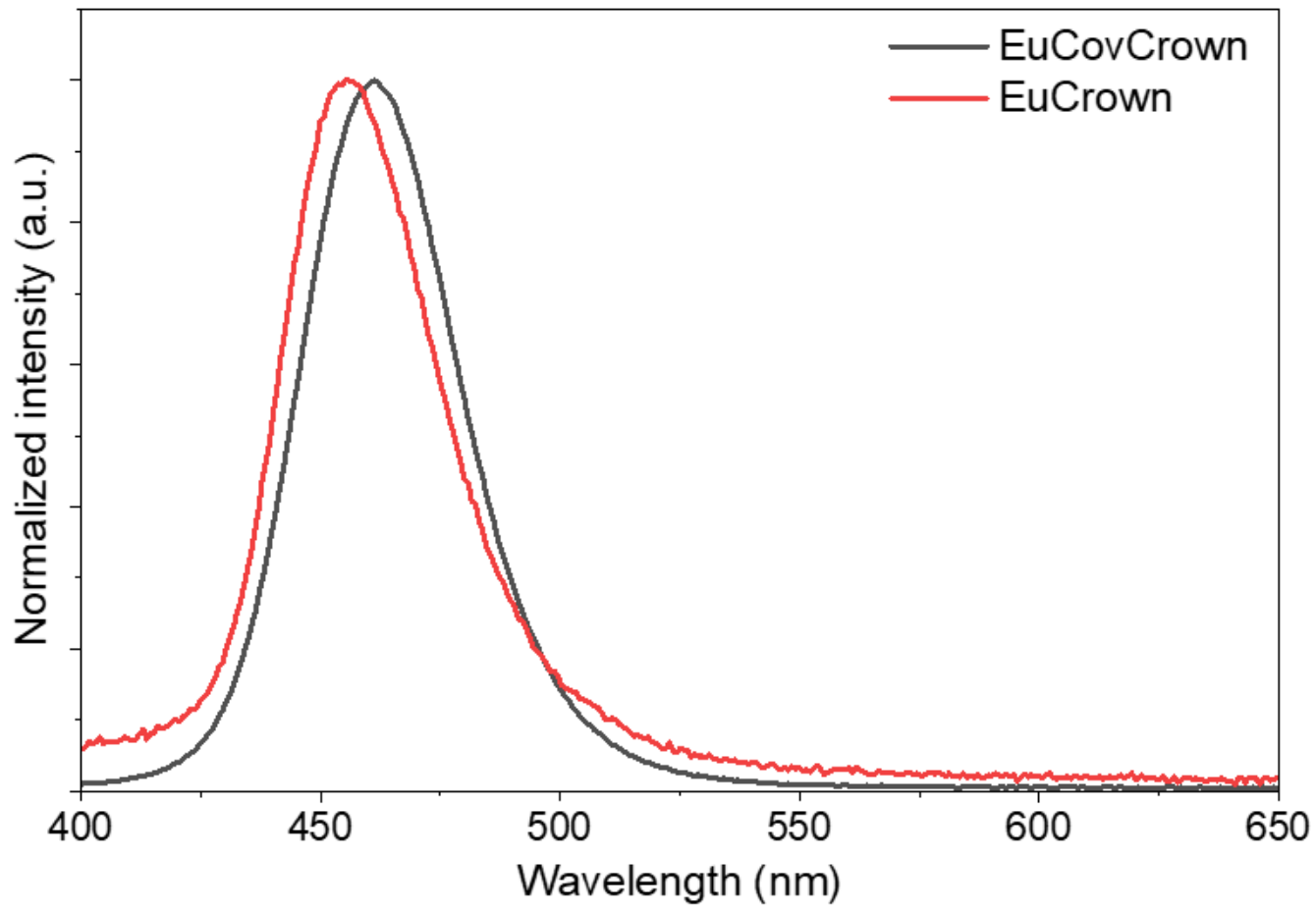

**Figure S7.** Emission spectra of EuCrown and EuCovCrown in dilute dispersions prepared in toluene/THF (3:1 v/v) upon excitation at 310 nm.

**Table S1.** Photophysical properties and derived decay rate constants of EuCrown and EuCovCrown measured in toluene/THF (3:1 v/v) suspension upon excitation at 310 nm. The radiative and nonradiative decay rate constants were calculated using $k_{\mathrm{r}} = \Phi_{\mathrm{PL}}/\tau$ and $k_{\mathrm{nr}} = (1 - \Phi_{\mathrm{PL}})/\tau$, where $\Phi_{\mathrm{PL}}$ is the photoluminescence quantum yield and $\tau$ is the excited-state lifetime.

| Emitter | PLQY (%) | $\tau$ (ns) | $k_{\mathrm{r}}$ ($\mathrm{s}^{-1}$) | $k_{\mathrm{nr}}$ ($\mathrm{s}^{-1}$) |
|---|---|---|---|---|
| EuCrown | 90 | 820 | $1.10 \times 10^6$ | $0.12 \times 10^6$ |
| EuCovCrown | 88 | 950 | $0.93 \times 10^6$ | $0.13 \times 10^6$ |

## S3.1 Ultraviolet Photoelectron Spectroscopy (UPS) of Eu(II) Emitters

UPS measurements were performed using a helium discharge lamp (UVS10/35) with the He-I excitation line at 21.22 eV. Photoelectrons were detected using a Phoibos 100 hemispherical analyzer equipped with an HSA3500 power supply. An acceleration voltage of $-$ 8 V was applied between the sample and analyzer to enhance electron collection. The base pressure of the chamber was $5 \times 10^{-10}$ mbar (rising to $\sim 1 \times 10^{-9}$ mbar during measurement). Samples were deposited on argon-sputtered gold foils and transferred under vacuum without ambient exposure. Energy calibration was performed using the Fermi edge of sputter-cleaned silver, yielding an instrumental resolution of 0.04 eV. Considering additional uncertainties (sample preparation, charging, and evaluation), the total estimated measurement uncertainty is approximately 70 meV. UPS measurements were performed on neat films of EuCrown and EuCovCrown to determine their ionization energies (**Figure S8**). The emitters were handled under inert conditions during preparation and loading into the deposition chamber. The extracted ionization energies are 6.26 eV for EuCrown and 5.89 eV for EuCovCrown, which qualitatively agree with the DFT results of 5.86 eV and 5.46 eV. The constant shift of $\approx$ 0.4 eV can be explained with the vertical nature of the UPS experiment (ionization without structural relaxation), whereas the DFT values are calculated assuming full relaxation. In fact, the vertical DFT ionization energies of 6.13 eV and 5.84 eV for EuCrown and EuCovCrown, respectively, are in good quantitative agreement with the UPS measurements.

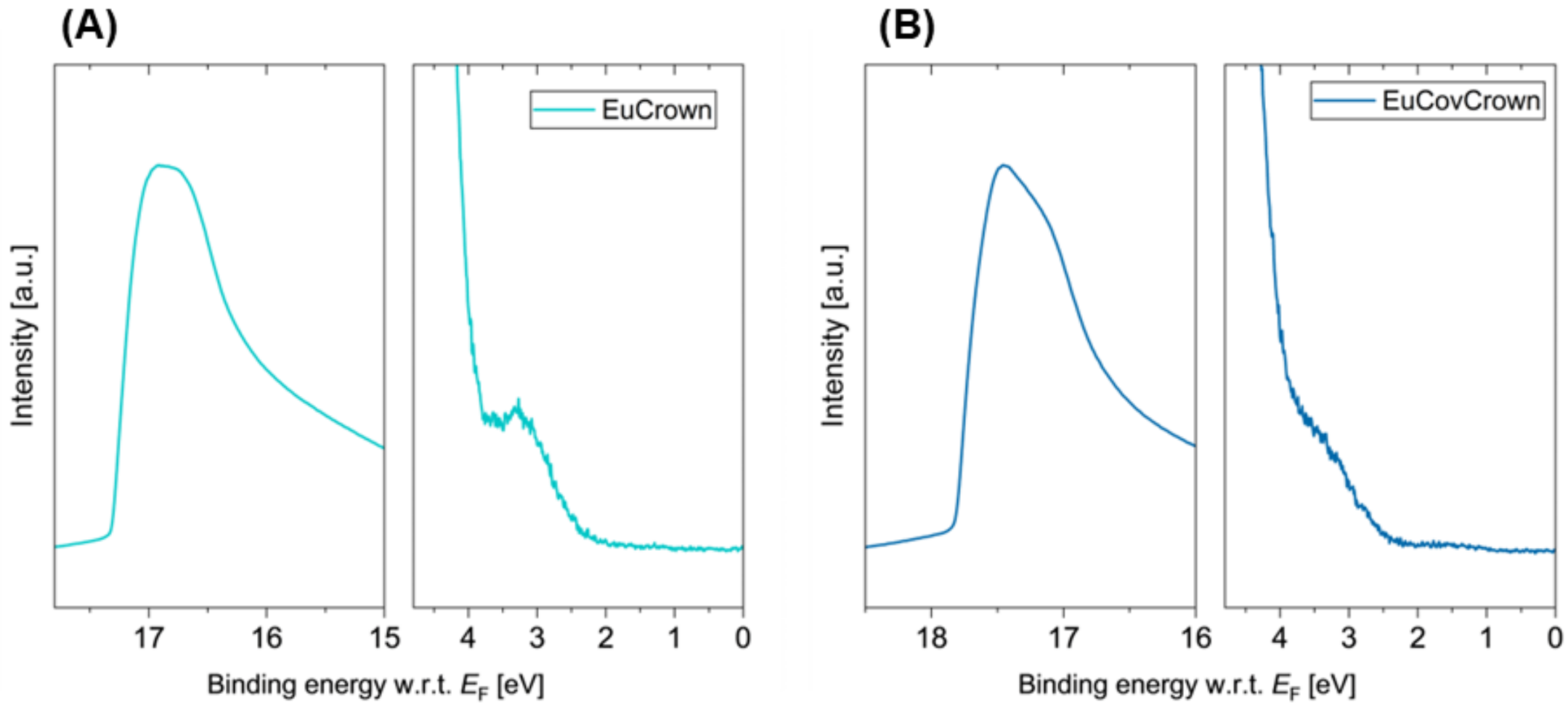

**Figure S8.** Ultraviolet photoelectron spectroscopy spectra of (A) EuCrown and (B) EuCovCrown. The secondary electron cutoff and HOMO onset were used to extract ionization energies (IE) and HOMO levels.

## S4 OLED Test Fabrication

To evaluate the processability of the emitters via vacuum deposition and examine their electroluminescent performance, bottom-emissive OLEDs were fabricated on pre-patterned indium tin oxide (ITO)-coated glass substrates (sheet resistance $\sim 15\ \Omega\ \mathrm{sq}^{-1}$). Prior to deposition, the substrates were cleaned. All organic and inorganic layers were deposited by thermal evaporation in a Kurt J. Lesker high-vacuum deposition system operated at base pressures below $1 \times 10^{-8}$ mbar. Deposition rates were monitored and controlled using individual quartz crystal microbalances (QCMs) and were typically set to 0.5–1.0 $\mathrm{Å\ s^{-1}}$ for organic layers, 0.2 $\mathrm{Å\ s^{-1}}$ for $MoO_3$, and 2 $\mathrm{Å\ s^{-1}}$ for the aluminum cathode. Following fabrication, the devices were transferred directly into a nitrogen-filled glovebox without exposure to ambient conditions and encapsulated using a glass cavity cap, a UV-curable epoxy, and a moisture/oxygen getter.

The chemical structures of the organic materials employed in the device stack are shown in **Figure S9**. These include SimCP2 as the hole-transport layer, SiDBFCz as the host matrix in the emissive layer, mSiTrz as the electron-transport layer, and TSPO1 doped with ytterbium as the electron-injection layer.

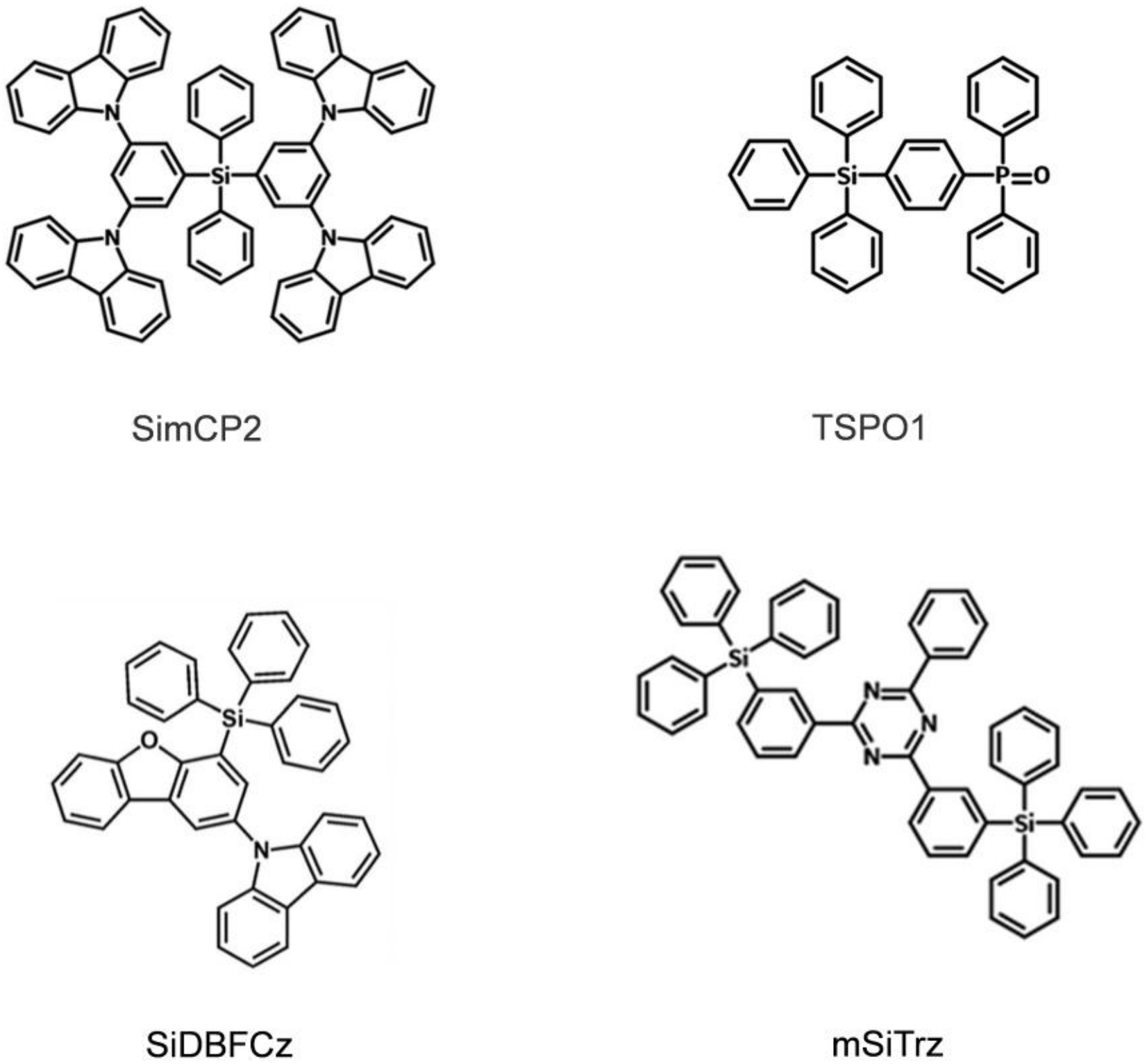

**Figure S9.** Chemical structures of the organic materials used in the OLED device stack.

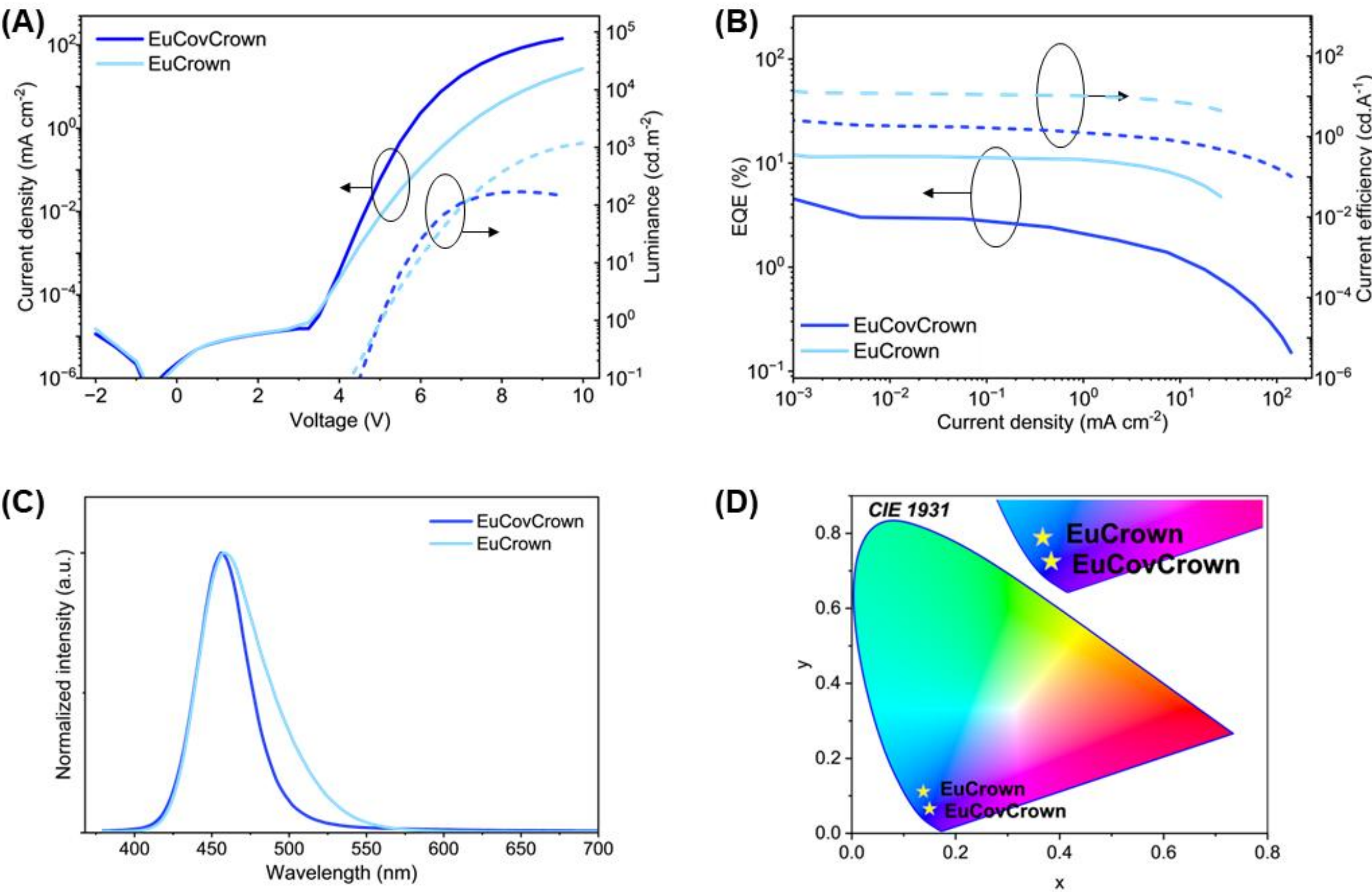

**Figure S10.** (A) Current density–voltage–luminance (*J–V–L*) characteristics of OLEDs incorporating EuCrown and EuCovCrown emitters. (B) Current efficiency and external quantum efficiency (EQE) as a function of current density. (C) Normalized electroluminescence spectra of OLEDs recorded at an operating voltage of 7 V. (D) CIE 1931 chromaticity coordinates of the corresponding devices.

**Table S2.** Electroluminescent characteristics of the OLEDs.

| Device | $V$ [V] | $EQE_{max}$ [%] | $CE_{max}$ [cd $A^{-1}$] | $\lambda_{max}$ [nm] | FWHM [nm] | CIE (x, y) |
|---|---|---|---|---|---|---|
| EuCrown | 3.3[a)] / 5.0[b)] | 12.3 | 12.8 | 458 | 50 | (0.14, 0.11) |
| EuCovCrown | 3.3[a)] / 5.0[b)] | 3.0 | 1.9 | 456 | 36 | (0.15, 0.06) |

$V$: [a)] voltage at the onset of a steep exponential increase in current density following the leakage-current regime, [b)] voltage at 1 cd $m^{-2}$; $EQE_{max}$: maximum external quantum efficiency; $CE_{max}$: maximum current efficiency; $\lambda_{max}$: peak electroluminescence wavelength; FWHM: full width at half maximum; CIE (x, y): 1931 color coordinates.

Bottom-emissive OLEDs were fabricated using a multilayer architecture using an ambipolar host suitable with respect to its excitonic energies, i.e., the triplet energy is higher than the emissive species of the emitters: ITO (Indium tin oxide) / $MoO_3$ (Molybdenum trioxide) [2 nm] / SimCP2 (Bis[3,5-di(9H -carbazol-9-yl)phenyl]diphenylsilane) [70 nm] / SiDBFCz (9-[4-(Triphenylsilyl)-2-dibenzofuranyl]-9H-carbazole):Emitter [30 wt%, 20 nm] / mSiTrz (2-Phenyl-4,6-bis(3-(triphenylsilyl)phenyl)-1,3,5-triazine) [5 nm] / TSPO1 (Diphenyl[4-(triphenylsilyl)phenyl]phosphine oxide):Yb (Ytterbium) [1 vol%, 25 nm] / Al [120 nm].

The current density–voltage–luminance (*J*–*V*–*L*) characteristics, external quantum efficiency, and electroluminescence properties are shown in **Figure S10**, and the key device parameters are summarized in **Table S2**. Both EuCrown- and EuCovCrown-based devices exhibit deep-blue electroluminescence with peak emission wavelengths at 458 and 456 nm, respectively. While the EL spectra are similar in shape, the EuCovCrown device shows a noticeably narrower FWHM of 36 nm compared to 50 nm for EuCrown, resulting in improved color purity, as reflected by the corresponding CIE 1931 chromaticity coordinates of (0.15, 0.06). Both devices display a turn-on voltage of approximately 3.3 V, defined by the onset of a steep exponential increase in current density following the leakage-current regime and reaching electroluminescence of 1 cd $m^{-2}$ at approximately 5.0 V. The EuCrown-based OLEDs reach a maximum EQE of 12.3% and a peak current efficiency of 12.8 cd $A^{-1}$, whereas the EuCovCrown devices show a significantly lower maximum EQE of 3.0% with a corresponding current efficiency of 1.9 cd $A^{-1}$. Both devices exhibit pronounced efficiency roll-off at higher current densities. The observed device performance is likely limited by charge injection and balance arising from energy level alignment, as well as changes in the intrinsic emission characteristics of the emitters within the emissive layer; nevertheless, these results demonstrate the processability of the materials via vacuum deposition and confirm their functionality as emitters in OLED applications.

## S5 Photoluminescence Studies of Host–Emitter System

### S5.1 Host Materials

Three representative OLED host materials—TAPC, SiDBFCz, and B3PyPB—were selected to systematically probe how europium(II) emitters respond to host environments with distinct electronic structure and coordinating character. These hosts are used commonly in OLED architectures and provide a controlled platform to study the energetic electron-confinement effects and specific chemical host–emitter interactions. TAPC is a hole-transporting host with a relatively shallow LUMO. Its tertiary amines are sterically embedded within the molecular framework, reducing the likelihood of direct coordination to the Eu(II) center. SiDBFCz represents an ambipolar host with intermediate LUMO depth and carbazole-based donor units, offering increased electron-accepting character while maintaining limited Lewis basicity toward the metal center. In contrast, B3PyPB is an electron-transporting host featuring pyridine moieties with exposed nitrogen atoms, which can act as strong Lewis bases and potential coordination sites for lanthanide cations.

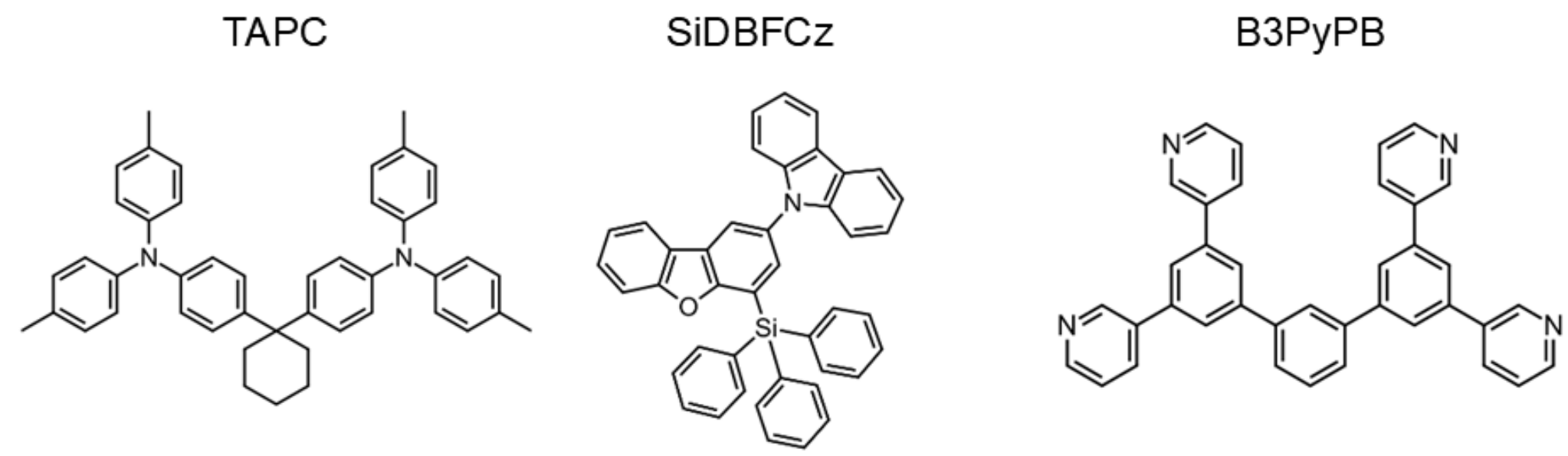

**Figure S11.** Chemical structures of the host materials employed in the photophysical study: TAPC, SiDBFCz, and B3PyPB.

Together, these hosts span a broad range of electron-accepting energies and coordinating strengths, enabling a systematic assessment of both excited-state electron confinement and specific chemical host–emitter interactions in Eu(II)-based systems.

## S5.2 Photophysical Properties of Host Materials

To characterize the optical properties of the host materials, thin films were prepared by thermal evaporation on quartz substrates. Absorption spectra were recorded using a Shimadzu UV–Vis spectrophotometer, while emission and excitation scans were acquired using a Horiba Fluoromax-4 spectrofluorometer inside a nitrogen-filled glovebox.

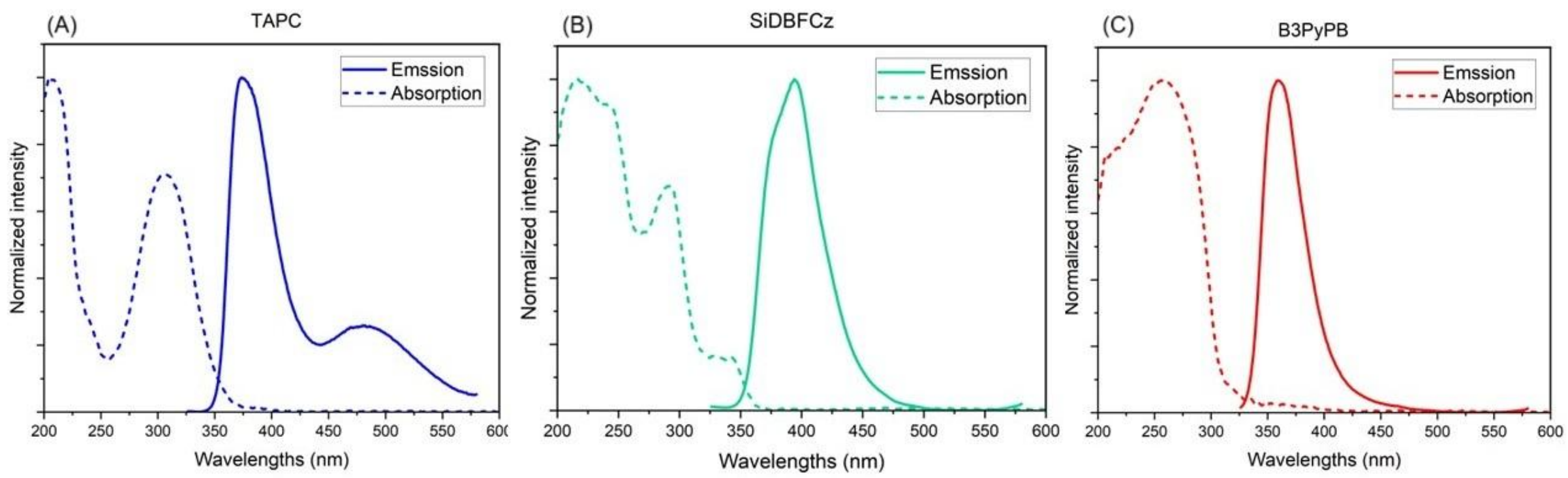

**Figure S12.** Absorption and emission spectra of the host materials: (A) TAPC, (B) SiDBFCz, and (C) B3PyPB, recorded from thin films deposited on quartz substrates. Emission spectra were obtained under 300 nm excitation.

To determine host triplet energies, low-temperature phosphorescence spectra were measured at 77 K from dilute host films (1 wt%) embedded in a PMMA matrix using 275 nm LED excitation.

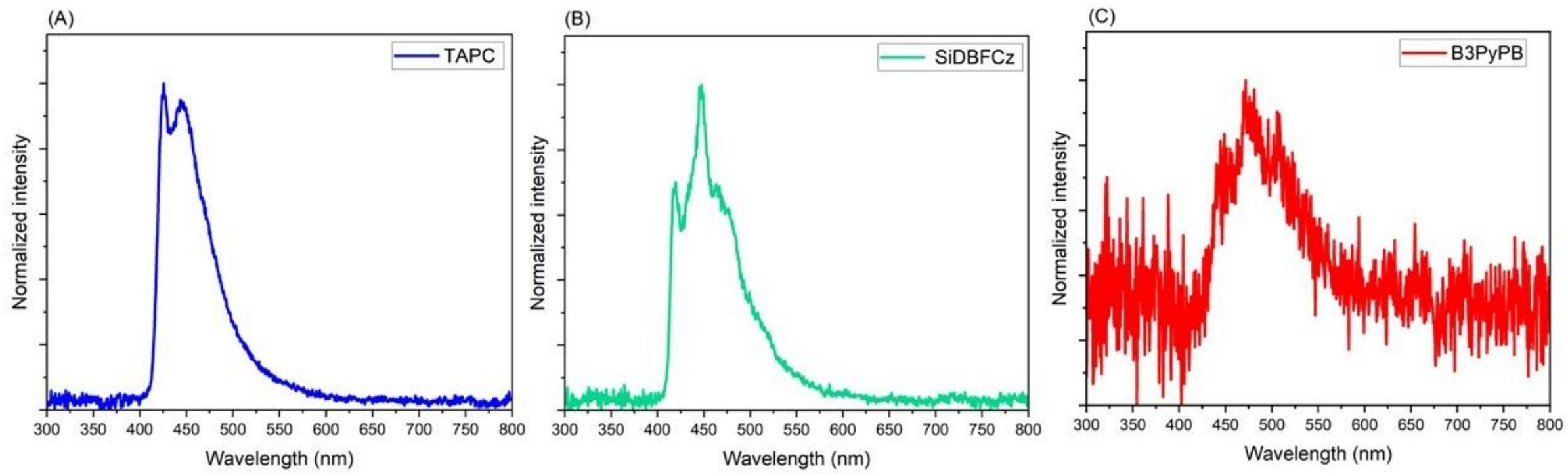

**Figure S13.** Low-temperature (77 K) phosphorescence spectra of the host materials: (A) TAPC, (B) SiDBFCz, and (C) B3PyPB, each doped at 1 wt% in a PMMA matrix. The onset of spectra was used to determine the host triplet energies ($T_1$).

## S5.3 Energy-Level of Host Materials

The frontier energy levels of the host materials were determined by combining ultraviolet photoelectron spectroscopy (UPS) and optical absorption measurements. Specifically, the highest occupied molecular orbital (HOMO) energies, corresponding to the ionization energies, were obtained directly from UPS, while the optical gaps ($E_{\mathrm{opt}}$) were extracted from the onset of the UV–Vis absorption spectra. The lowest unoccupied molecular orbital (LUMO) energies were then derived from the HOMO energies and optical gaps according to:

$$E_{\mathrm{LUMO}} = E_{\mathrm{HOMO}} + E_{\mathrm{opt}}, \qquad (1)$$

In addition, triplet energies ($T_1$) were obtained from the low-temperature phosphorescence measurements described in section S5.2, ensuring that triplet-energy transfer from the host to the Eu(II) emitters is energetically unfavorable and does not account for emission quenching.

### S5.3.1 *Ultraviolet Photoelectron Spectroscopy (UPS) of Hosts*

UPS measurements were carried out using the same experimental setup and analysis procedure as described for the emitter materials. A helium discharge lamp (He I, 21.22 eV) and a hemispherical analyzer were employed under high-vacuum conditions (~$10^{-10}$ mbar). Samples were deposited on argon-sputtered gold substrates and transferred without exposure to ambient conditions. Energy calibration was performed using the Fermi edge of sputter-cleaned silver, yielding an instrumental resolution of 0.04 eV. Taking into account additional uncertainties arising from sample preparation, charging effects, and data evaluation, the overall measurement uncertainty is estimated to be approximately 70 meV.

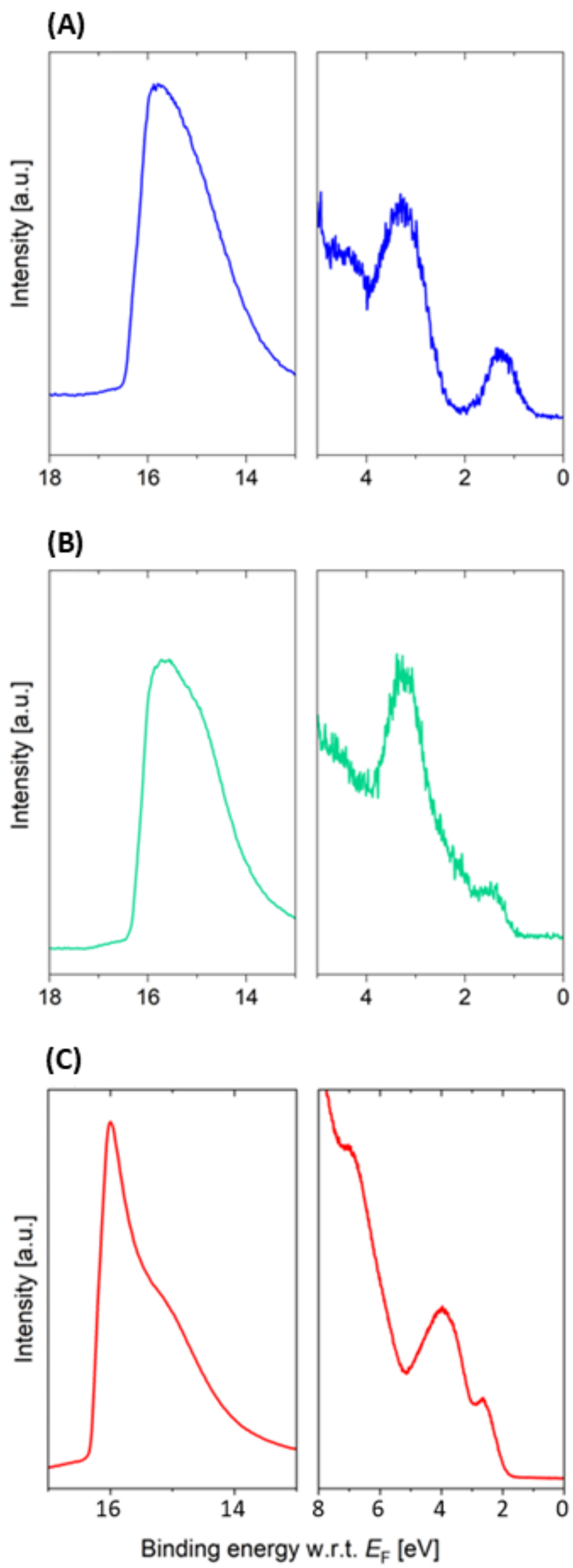

**Figure S14.** UPS spectra of the host materials: (A) TAPC, (B) SiDBFCz, and (C) B3PyPB. The secondary electron cutoff and HOMO onset were used to extract ionization energies (IE) and HOMO levels.

Ionization energies were determined from:

$$\mathrm{IE} = h\nu - E_{\mathrm{cutoff}} + E_{\mathrm{HOMO}}, \quad (2)$$

where $h\nu = 21.22$ eV, $E_{\mathrm{cutoff}}$ is the secondary electron cutoff, and $E_{\mathrm{HOMO}}$ is the HOMO onset relative to the Fermi level. A summary of the experimentally determined host energy levels is given in **Table S3**. The LUMO values reported here are used in the main text to compare host electron-accepting energies with the excited-state ionization energies (ES–IE) of the Eu(II) emitters.

**Table S3.** Experimentally determined energy levels of OLED host materials. HOMO energies were obtained from UPS, optical gaps from UV–Vis absorption, and LUMO energies were derived via Eq. 1. Triplet energies were determined from low-temperature phosphorescence measurements. All values are given in eV.

| Host | HOMO | $E_{\mathrm{opt}}$ | LUMO | $T_1$ |
|---|---|---|---|---|
| TAPC | −5.4 | 3.5 | −2.0 | 3.0 |
| SiDBFCz | −6.0 | 3.4 | −2.5 | 3.0 |
| B3PyPB | −6.8 | 4.0 | −2.8 | 2.9 |

## S5.4 Sample Design and Fabrication for Photoluminescence Studies

For the photoluminescence study, samples were prepared on cleaned quartz substrates using thermal evaporation in a Kurt J. Lesker high-vacuum deposition system operated at base pressures below $1 \times 10^{-8}$ mbar. Material deposition was carried out under controlled conditions, with rates ranging from 0.1 to 2 Å/s, depending on the specific doping concentration required for each sample. To enhance light absorption at lower emitter concentrations, the thickness of the active layer was adjusted. The schematic of the sample design is shown in **Figure S15**. The sample sets for each emitter were prepared as described in

**Table S4**, where the numbers in parentheses denote the thickness of each layer in nanometers (nm).

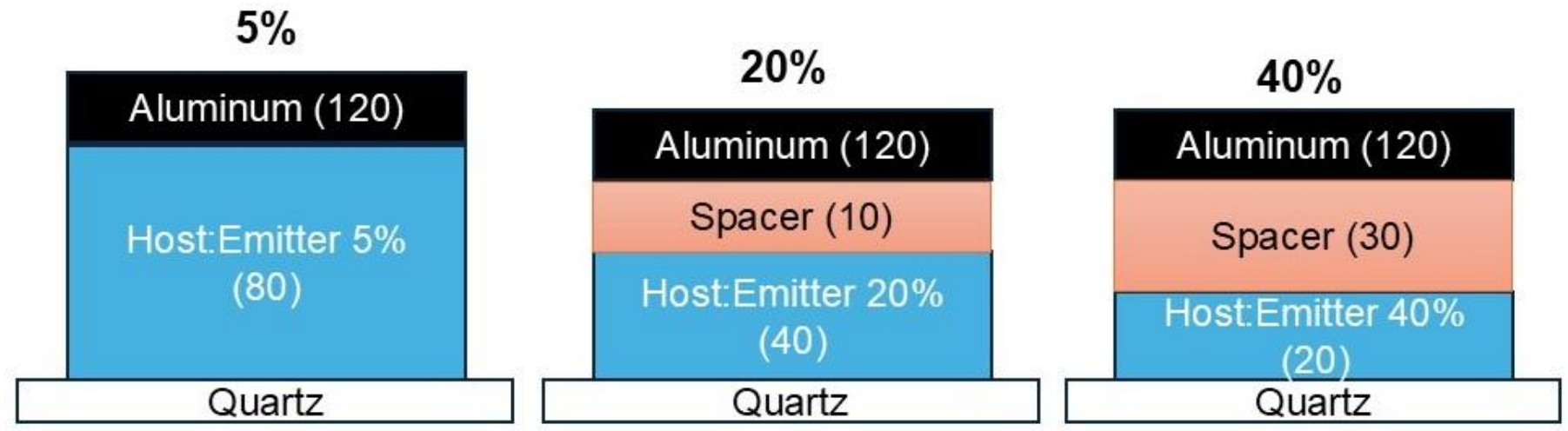

**Figure S15.** Schematic sample architectures for photophysical studies at different doping levels. Samples were fabricated on quartz substrates with layer thicknesses (nm) given in parentheses, using hosts (TAPC, SiDBFCz, B3PyPB) and Eu-complexes (EuCrypt, EuCrown, EuCovCrown) as described in

**Table S4.** Sample architectures of EuCrypt, EuCrown, and EuCovCrown in different hosts and doping levels. Layer thicknesses are given in nanometers (nm).

| Sample | Structure | Emitter |
|---|---|---|
| | EuCrypt | |
| S11 | Quartz/TAPC:5%EuCrypt (80)/Al (120) | EuCrypt |
| S12 | Quartz/TAPC:20%EuCrypt (40)/2SiDBF (10)/Al (120) | EuCrypt |
| S13 | Quartz/TAPC:40%EuCrypt (20)/2SiDBF (30)/Al (120) | EuCrypt |
| S21 | Quartz/SiDBFCz:5%EuCrypt (80)/Al (120) | EuCrypt |
| S22 | Quartz/SiDBFCz:20%EuCrypt (40)/2SiDBF (10)/Al (120) | EuCrypt |
| S23 | Quartz/SiDBFCz:40%EuCrypt (20)/2SiDBF (30)/Al (120) | EuCrypt |
| S31 | Quartz/B3PyPB:5%EuCrypt (80)/Al (120) | EuCrypt |
| S32 | Quartz/B3PyPB:20%EuCrypt (40)/2SiDBF (10)/Al (120) | EuCrypt |
| S33 | Quartz/B3PyPB:40%EuCrypt (20)/2SiDBF (30)/Al (120) | EuCrypt |
| | EuCrown | |
| S11 | Quartz/TAPC:5%EuCrown (80)/Al (120) | EuCrown |
| S12 | Quartz/TAPC:20%EuCrown (40)/2SiDBF (10)/Al (120) | EuCrown |
| S13 | Quartz/TAPC:40%EuCrown (20)/2SiDBF (30)/Al (120) | EuCrown |
| S21 | Quartz/SiDBFCz:5%EuCrown (80)/Al (120) | EuCrown |
| S22 | Quartz/SiDBFCz:20%EuCrown (40)/2SiDBF (10)/Al (120) | EuCrown |
| S23 | Quartz/SiDBFCz:40%EuCrown (20)/2SiDBF (30)/Al (120) | EuCrown |
| S31 | Quartz/B3PyPB:5%EuCrown (80)/Al (120) | EuCrown |
| S32 | Quartz/B3PyPB:20%EuCrown (40)/2SiDBF (10)/Al (120) | EuCrown |
| S33 | Quartz/B3PyPB:40%EuCrown (20)/2SiDBF (30)/Al (120) | EuCrown |
| | EuCovCrown | |
| S11 | Quartz/TAPC:5%EuCovCrown (80)/Al (120) | EuCovCrown |
| S12 | Quartz/TAPC:20%EuCovCrown (40)/2SiDBF (10)/Al (120) | EuCovCrown |
| S13 | Quartz/TAPC:40%EuCovCrown (20)/2SiDBF (30)/Al (120) | EuCovCrown |
| S21 | Quartz/SiDBFCz:5%EuCovCrown (80)/Al (120) | EuCovCrown |
| S22 | Quartz/SiDBFCz:20%EuCovCrown (40)/2SiDBF (10)/Al (120) | EuCovCrown |
| S23 | Quartz/SiDBFCz:40%EuCovCrown (20)/2SiDBF (30)/Al (120) | EuCovCrown |
| S31 | Quartz/B3PyPB:5%EuCovCrown (80)/Al (120) | EuCovCrown |
| S32 | Quartz/B3PyPB:20%EuCovCrown (40)/2SiDBF (10)/Al (120) | EuCovCrown |
| S33 | Quartz/B3PyPB:40%EuCovCrown (20)/2SiDBF (30)/Al (120) | EuCovCrown |

## S5.5 Considerations for the Evaluation of Photoluminescence Studies

Photoluminescence quantum yield (PLQY) is a common method for evaluating emissive materials for organic light-emitting diodes (OLEDs), as it reflects the balance between radiative and non-radiative decay of the excited state. For Eu(II) complexes, however, PLQY measurement methods commonly used for organic emitters are not straightforward to apply. Measurements in dilute solution are not fully suitable because the Eu(II) complexes have limited solubility and may also undergo partial solvolysis and/or exhibit emission characteristics that differ from those of the solid-state complexes, as shown previously.[1] Therefore, to assess the intrinsic radiative efficiency of the Eu(II) emitters while minimizing intermolecular interactions, PLQY values of the emitters were determined using dilute dispersions of well-dispersed particles. These dispersion-based values provide a reference for the emissive properties of the Eu(II) cores, but they do not directly predict the behavior of the emitters in thin-film host–guest environments.

PLQY determination in a thin film host–guest system is further complicated for several reasons. The main challenge is the strong spectral overlap between the excitation bands of Eu(II) complexes and those of common organic host materials. As a result, selective excitation of the emitter without simultaneous host excitation is not feasible, and energy-transfer processes from the host to the guest would contribute to the measured signal, leading to ambiguous PLQY values that reflect system properties rather than intrinsic emitter characteristics. This is particularly relevant for parity-allowed 4f–5d transitions, which exhibit relatively low oscillator strengths compared to π–π* transitions in organic hosts. In addition, the emission in thin films is highly sensitive to air exposure and is rapidly reduced upon exposure to the ambient atmosphere.

To ensure measurement stability, the films were encapsulated with an additional organic layer followed by an aluminum capping layer, as shown in the sample design section. While this encapsulation preserves the emissive properties of the films, it makes the samples reflective on one side, such that conventional integrating-sphere-based PLQY measurements are no longer valid due to directional reflection effects.[2,3]

Hence, instead of relying on thin-film PLQY measurements to evaluate emission efficiency, which would be strongly influenced by host excitation and optical-collection effects, we assess emitter performance through excited-state lifetime analysis, which provides a more direct probe of host-induced non-radiative decay processes. The radiative emission efficiency ($\eta_{\mathrm{em}}$) can be expressed as:

$$\eta_{\mathrm{em}} = \frac{k_{\mathrm{r}}}{k_{\mathrm{r}}+k_{\mathrm{nr}}}, \qquad (3)$$

where $k_{\mathrm{r}}$ and $k_{\mathrm{nr}}$ denote the radiative and non-radiative recombination rate constants, respectively. For Eu(II) emitters, the localized nature of the $4f$–$5d$ excited state implies that $k_{\mathrm{r}}$ is largely an intrinsic property of the complex, whereas variations in $k_{\mathrm{nr}}$ predominantly originate from extrinsic interactions with the surrounding host environment. Consequently, when the same emitter is embedded in different host materials under otherwise identical conditions, changes in the excited-state lifetime can be directly attributed to host-induced variations in non-radiative decay pathways arising from host and emitter interactions. Well-established time-correlated single-photon counting (TCSPC) measurements, therefore, provide a quantitative approach to probe relative changes in the excited-state decay dynamics of the Eu(II) complexes as a function of the selected host material. These changes can be related to variations in the radiative emission efficiency, which contributes to the photoluminescence quantum yield according to $\mathrm{PLQY} = \eta_{\mathrm{abs}}\eta_{\mathrm{ISC}}\eta_{\mathrm{em}}$, where $\eta_{\mathrm{abs}}$, $\eta_{\mathrm{ISC}}$, and $\eta_{\mathrm{em}}$ denote the absorption, intersystem crossing, and emissive efficiencies, respectively.[4,5]

## S5.6 Photoluminescent Study of Host and Emitter Blend

Steady-state spectra obtained with an excitation wavelength of 300 nm, together with transient decay measurements recorded using the monochromator set at the peak emission of the emitters in the host matrix, are presented in **Figure S16-S18**, corresponding to EuCrypt, EuCrown and EuCovCrown at different concentrations within various host materials, respectively.

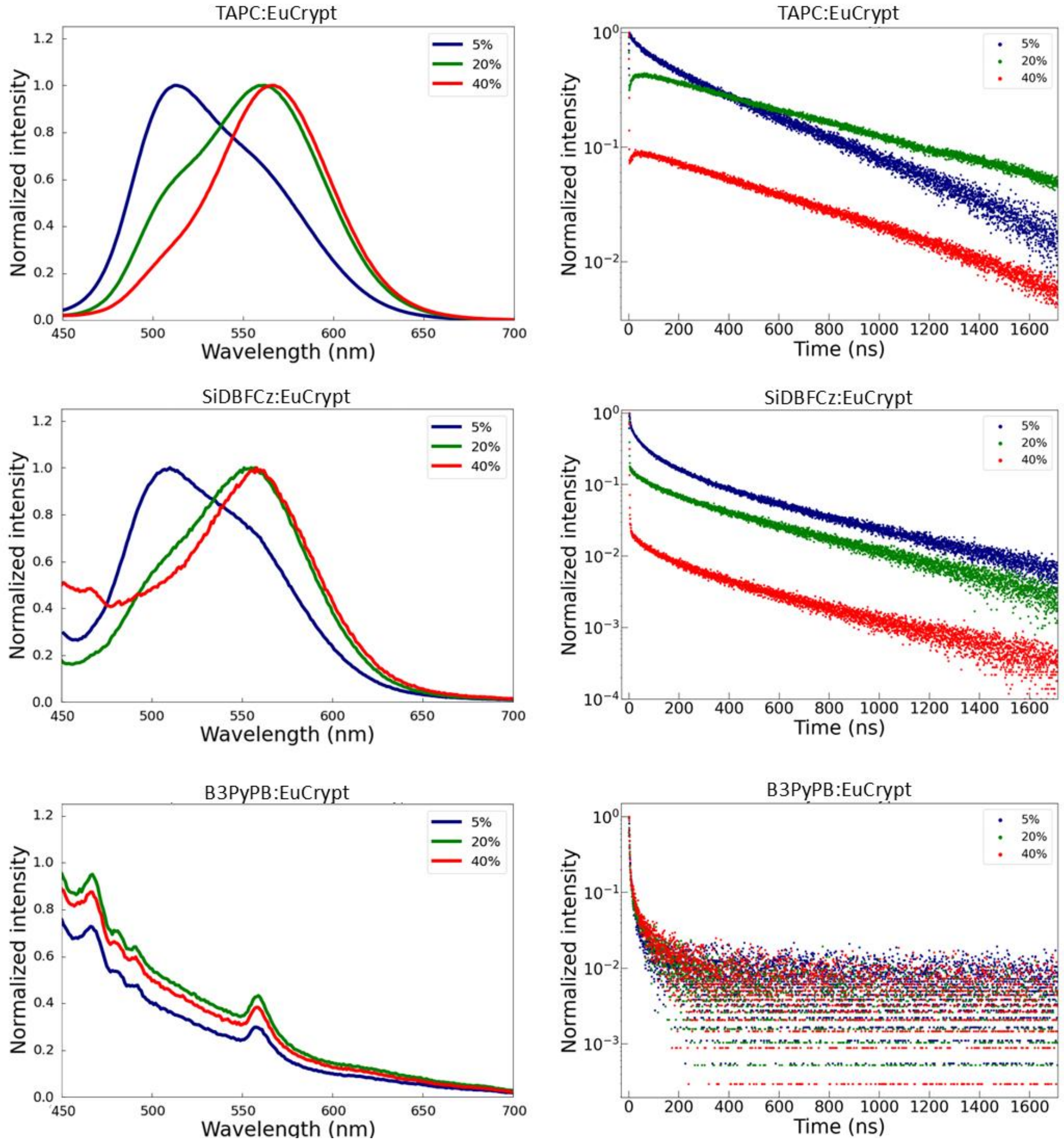

**Figure S16.** Steady-state spectra and transient decay of EuCrypt with different concentrations within different hosts with 300 nm excitation wavelength.

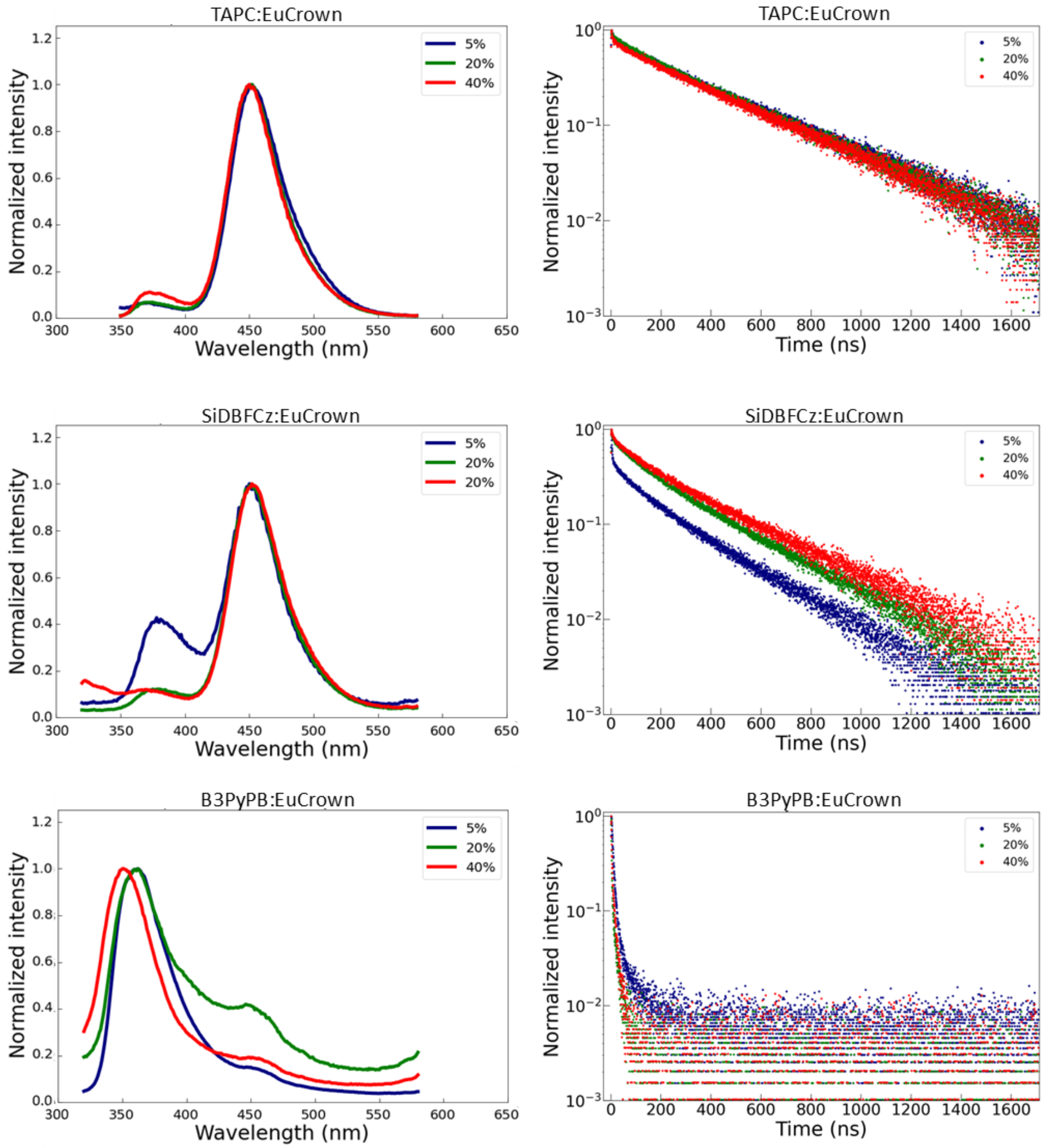

**Figure S17.** Steady-state spectra and transient decay of EuCrown with different concentrations within different hosts with 300 nm excitation wavelength.

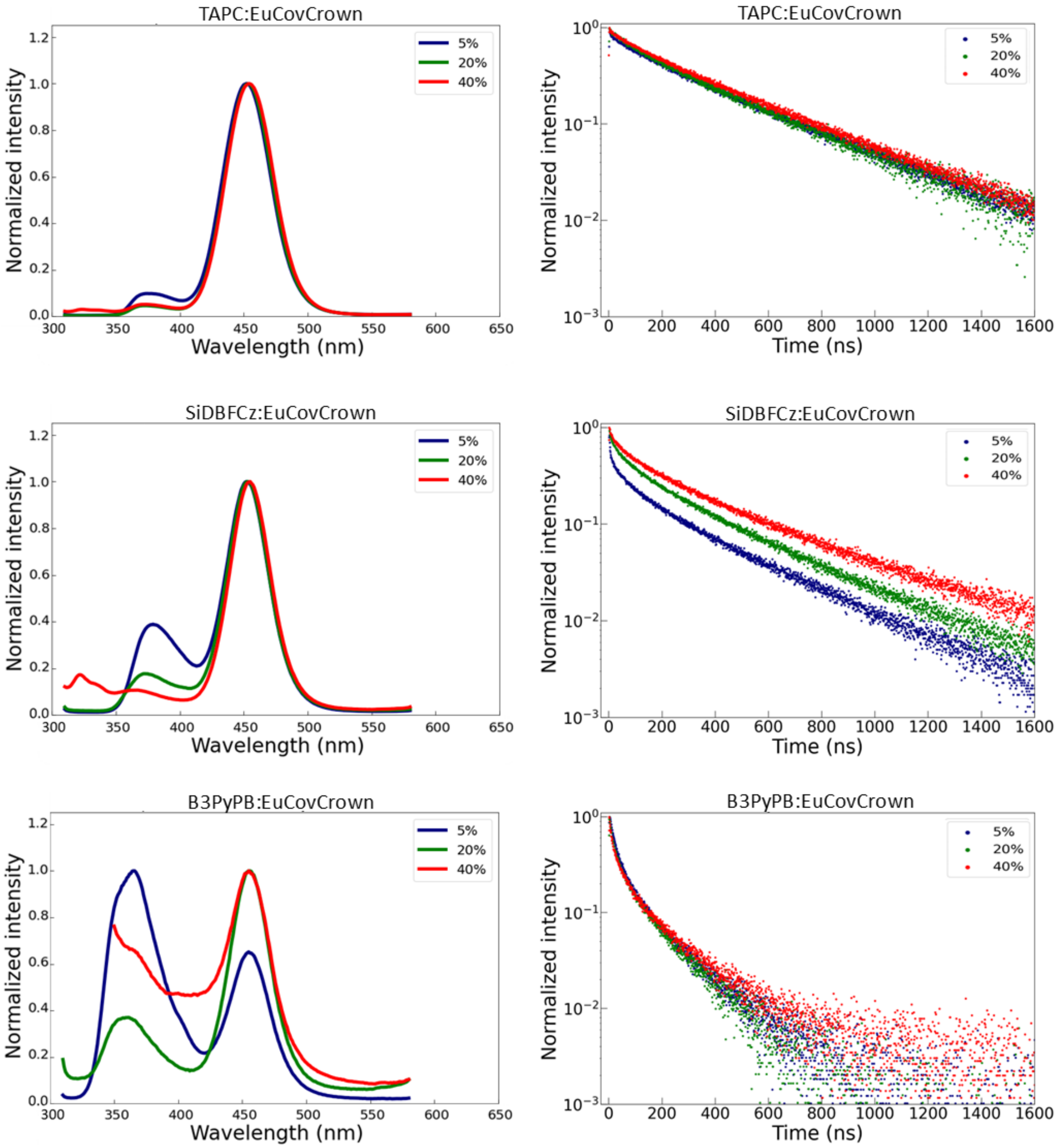

**Figure S18.** Steady-state spectra and transient decay of EuCovCrown with different concentrations within different hosts with 300 nm excitation wavelength.

## S5.7 Gaussian Fitting of Emission Spectra

To analyze the emission shape of the Eu complexes, we performed peak fittings using OriginLab Pro. The aim was to assess whether the emission band can be described by a symmetric single Gaussian function or whether an asymmetric (bi-Gaussian) function is required.

For EuCovCrown, the emission band is symmetric, and the data are perfectly described by a single Gaussian, highlighting that this emitter exhibits a clean Gaussian emission profile. The function used was:

$$y(x) = y_0 + \frac{A}{w\sqrt{\pi/(4\ln 2)}} \exp\left[-4\ln(2)\,\frac{(x-x_c)^2}{w^2}\right], \qquad (4)$$

where $y_0$ is the baseline offset, $x_c$ is the peak position, $A$ is the integrated peak area, and $w$ is the full width at half maximum (FWHM).

In contrast, the emission of EuCrown shows a clear asymmetry. In this case, the spectra are best fitted with a bi-Gaussian (asymmetric Gaussian) function:

$$y(x) = \begin{cases} y_0 + H\exp\left[-\frac{(x-x_c)^2}{2w_1^2}\right], & x < x_c, \\ y_0 + H\exp\left[-\frac{(x-x_c)^2}{2w_2^2}\right], & x \geq x_c, \end{cases} \qquad (5)$$

where $H$ is the peak height, and $w_1$ and $w_2$ are the Gaussian widths.

The fitting results show excellent agreement between model and experimental spectra ($R^2 > 0.997$ in all cases). Residuals confirm that a single Gaussian function provides an excellent and nearly ideal description of EuCovCrown emission, whereas the emission of EuCrown requires an asymmetric bi-Gaussian fit to capture the deviation from symmetry. The Gaussian fits are shown in **Figure S19**, and the corresponding fitting parameters are listed in **Table S5**.

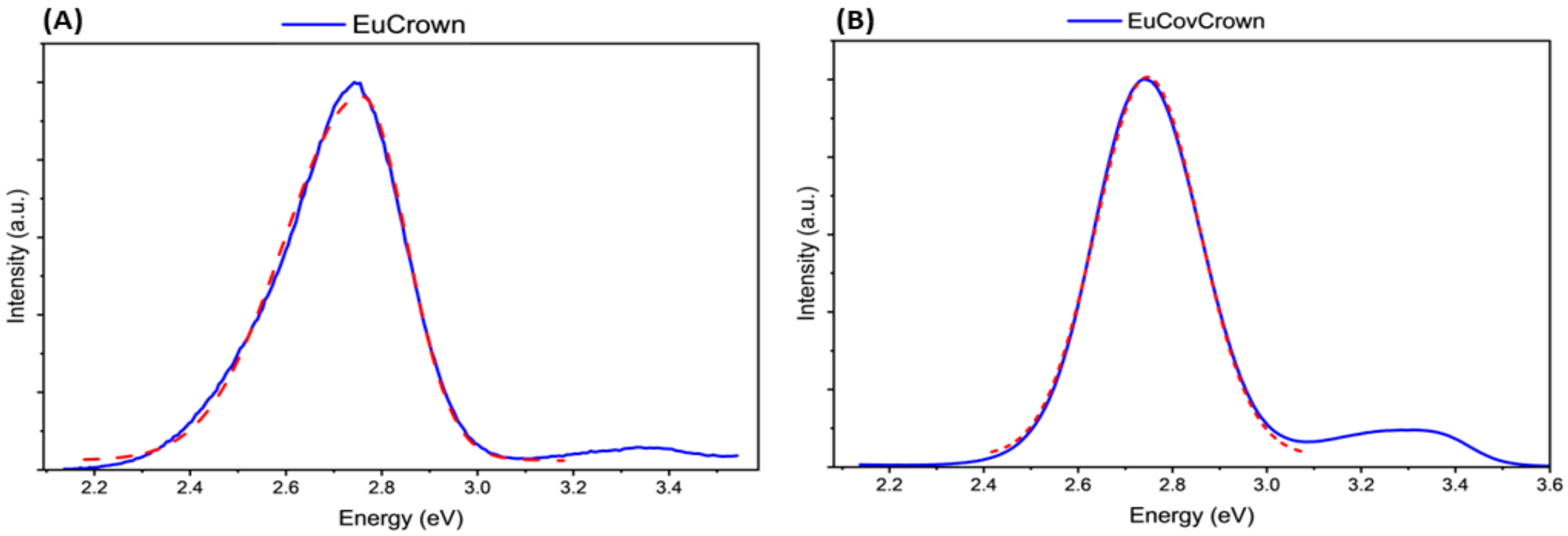

**Figure S19.** Gaussian fitting of the emission spectra of Eu complexes (5 wt% emitter doped in a TAPC host). (A) EuCrown emission spectrum fitted with a bi-Gaussian (asymmetric Gaussian) function, showing different left and right widths. (B) EuCovCrown emission spectrum fitted with a single Gaussian, highlighting its clean and symmetric profile. The excellent agreement between the experimental data and fitted curves ($R^2 > 0.997$) demonstrates that EuCovCrown exhibits a single-Gaussian emission profile, whereas EuCrown requires an asymmetric bi-Gaussian description.

**Table S5.** Gaussian fit parameters for EuCrown and EuCovCrown (5 wt% emitter in TAPC). Energies and widths are reported in eV. Note: In the single-Gaussian model, w corresponds to the full width at half maximum (FWHM). In the bi-Gaussian model, $w_1$ and $w_2$ are standard deviations ($\sigma$), which were converted to FWHM via FWHM = 2.35 $\sigma$.

| Emitter | Model | Peak center $x_c$ | FWHM | $R^2$ |
|---|---|---|---|---|
| EuCovCrown | Single Gaussian | 2.75 | 0.26 | 0.999 |
| EuCrown (left) | Bi-Gaussian ($w_1$) | 2.76 | 0.38 | 0.997 |
| EuCrown (right) | Bi-Gaussian ($w_2$) | 2.76 | 0.22 | |

## S6 Computational Details

All IE and ES-IE energies presented in this work include geometric relaxation, i.e., these are energy differences between the fully optimized Eu(II) and optimized Eu(III) structures for the IE, and $(4f - 5d)$ excited-state optimized Eu(II) and optimized Eu(III) for ES−IE. Note that obtaining the IE energy via the difference of the DFT energies of the ground state to that of the (relaxed) cationic complex comprising Eu(III) is essentially a more rigorous way than the common approach of taking the HOMO energy (Kohn-Sham eigenvalue) from a DFT/B3LYP calculation. The main advantage is that it largely eliminates a strong dependency on the employed density-functional, specifically the amount of Fock exchange in the functional, making results obtained with different functionals comparable. An additional advantage is that this more rigorous approach allows to account for the geometric relaxation of the cation, which is typically larger for the oxidation of a metal center in a complex, than for the oxidation of an organic molecule. These theoretical considerations are confirmed by applying them to the organic host materials, which produce HOMO (−IE) values in good agreement with UPS measurements (*cf.* **Table S3**).

Finally, we want to mention that in addition to the effective-core potential (ECP) calculations in the scalar relativistic approximation, we also conducted all-electron single point calculations using the more complete Douglas-Kroll-Hess Hamiltonian of second order with a point-nucleus approximation as described by Reiher and Wolf.[6] Albeit less approximate, the all-electron results are in worse agreement with the experimental emission and absorption energies, exhibiting a systematic red-shift of about 0.4 eV, while the calculated (ground state) IEs are smaller by the same amount. Together, these effects cancel, such that the ES−IE, which are the key quantity regarding electron-confinement, are very similar in the scalar-relativistic and all-electron calculations. Therefore, we decided to use the scalar-relativistic results throughout this work.

### S6.1 DFT Analysis of Dipole and Emission Energies

The substantial dipole moment of EuCrypt (12.7 D) arises from the asymmetric charge distribution caused by the two iodides being located on one side of the europium center. In the solid state, such a large dipole is consistent with strong electrostatic interactions and thus with limited volatility and salt-like character of the material. The calculated emission energy for the lowest conformer is 2.68 eV, which is higher than the experimental value reported by Li et al. (560 nm, 2.21 eV).[7] However, the emission energy is sensitive to small structural changes: already for the second conformer, in which both iodides coordinate the NH protons (+0.2 eV in the ground state), the predicted emission shifts to 2.38 eV, in better agreement with both the experimental emission energy and the Eu-I distances reported by Basal and coworkers.[8] Overall, these differences are within the usual error bars of DFT and may additionally reflect packing effects in the solid.

## S6.2 Calculating IE, ES-IE, and Ems in the Presence of Iodide

The central issue here is that the (ES−)IEs presented in this work include the effects of geometric relaxation, just like the emission energy. However, with iodide present in these structures at a fixed Eu-I distance, geometric relaxation cannot be easily included. Therefore, we started by calculating vertical IEs and absorption energies for fixed geometries from the scans. Subsequently, to put these into the context of the undisturbed emitters, the differences between the structures without iodide (point ∞) and the outside/inside adducts are calculated and added to the values calculated for the isolated emitters (first row in each table). Finally, we want to stress that since an anionic host would be less compact (further away from Eu) and have a more delocalized charge distribution than iodide, both of which reduce the electrostatic impact on Eu, the shifts between the isolated species, outside, and inside-adducts should perhaps be regarded as a lower bound that overestimates the influence of the host environment.